\documentclass[journal]{new-aiaa}

\usepackage{amssymb}
\usepackage{amsmath,amsfonts}
\usepackage{multirow}
\usepackage{hyperref}
\usepackage{lineno}
\usepackage{hyperref}
\usepackage{graphicx}
\usepackage{subcaption}
\usepackage{bm}
\usepackage{dcolumn}
\usepackage{booktabs}
\usepackage{lineno}
\usepackage{empheq}
\usepackage{mathtools}
\usepackage{booktabs}
\usepackage{makecell}
\usepackage{cellspace}
\usepackage{caption}
\usepackage{xcolor}
\usepackage{scalerel,stackengine}
\usepackage{xspace}
\usepackage[version=4]{mhchem}
\usepackage{siunitx}
\usepackage{tikz,pgfplots,pgfplotstable}
\usepackage{subcaption}
\usepackage[mathscr]{euscript}

\DeclareMathAlphabet{\mathcalligra}{T1}{calligra}{m}{n}

\newcommand{\red}[1]{{\textcolor{black}{#1}}}
\newcommand{\blue}[1]{{\textcolor{black}{#1}}}

\newcommand{\beq}{\begin{equation}}
\newcommand{\eeq}{\end{equation}}

\makeatletter
\def\ps@pprintTitle{%
\let\@oddhead\@empty
\let\@evenhead\@empty
\def\@oddfoot{\reset@font\hfil\thepage\hfil}
\let\@evenfoot\@oddfoot
}
\makeatother

\title{Modal analysis of oblique shock-induced flow dynamics in a supersonic reacting shear layer}

\author{Radouan Boukharfane\footnote{radouan.boukharfane@um6p.ma}}
\affil{Mohammed VI Polytechnic University (UM6P), College of Computing, Benguerir, Morocco}

\begin{document}
\date\today
\maketitle

\begin{abstract}
Efficient mixing in high-speed compressible flows, crucial for scramjet operation, can be significantly enhanced by shock wave interactions. This study employs Direct Numerical Simulations (DNS) to comprehensively examine the interaction between an oblique shock and a spatially developing turbulent mixing layer, contrasting inert and reacting (hydrogen-air combustion) cases. Utilizing streaming Dynamic Mode Decomposition (sDMD), we analyze four configurations: inert and reacting shear layers, both with and without shock impingement (at $\mathrm{Ma}_c = 0.48$). We evaluate the temporal mode growth rates, the evolution of vorticity thickness, and the spatial structures of dominant DMD modes to elucidate how shocks and heat release synergistically influence flow stability, mixing, and the underlying coherent dynamics. Results reveal that the oblique shock significantly amplifies Kelvin-Helmholtz instabilities, excites a broader spectrum of unstable temporal modes, and accelerates the growth of the vorticity thickness. Combustion-induced heat release further modifies this response, leading to a redistribution of energy among the DMD modes and indicating a complex coupled effect with shock dynamics, particularly in the enhanced excitation of high-frequency modes and the alteration of spatial structures. The modal analysis identifies distinct frequency bands associated with shock and combustion effects and characterizes the dominant spatial patterns, offering refined insights for controlling and enhancing mixing in high-speed propulsion flows.
\end{abstract}

\section{Introduction}

Turbulence, shock waves, and combustion interaction continues to be among the most fundamental and challenging problems in the field of fluid dynamics. Compressible and reactive flows display an overwhelming array of multi-scale, nonlinear interactions that influence transport, mixing, noise generation, and flame behavior in engineering applications such as supersonic jets, scramjets, and propulsion systems. These flows are characterized by inhomogeneous effects of shock–turbulence interaction, shock–flame interaction, and baroclinic generation of vorticity, which all exert strong influence on the global evolution and the coherent structures of the system.
Recent developments in high-performance computing and numerical techniques have enabled large-eddy simulation (LES) and direct numerical simulation (DNS) to become invaluable tools for investigating compressible and reactive flows at high resolution. These state-of-the-art techniques grant access to the large-scale spatiotemporal development of turbulence and allow for close examinations of the interactions among vortical structures, compressibility effects, shock dynamics, and heat release phenomena. For instance, DNS studies of compressible jets and mixing layers have shed light on the intricate dynamics induced by shock–turbulence interactions \citep{pirozzoli2011turbulence,boukharfane2018evolution} and acoustic and entropy wave amplifications \citep{mahesh2013interaction}.
In reactive flows, turbulence-combustion interaction adds another level of complexity since exothermic reaction changes local flow topology and either amplifies or quenches instabilities. Initial studies \citep{mathew2008effects} already demonstrated that heat release and compressibility both together affect the spectral characteristic and geometry of turbulence with consequences for noise generation and mixing efficiency. Further, shock–flame interaction studies \citep{fureby2009large} have revealed phenomena like flame acceleration, local quenching, and baroclinic generation of vorticity, which are all highly thermodynamic state and shock geometry sensitive.
In spite of these results, the knowledge of such interactions is still incomplete, particularly in cases where several nonlinear processes—like shock compression, turbulent straining, acoustic feedback, and chemical heat release—occur together. These complexities demand ongoing reliance on high-fidelity simulations, which are not only utilized to investigate open issues but also to create datasets that can be reduced-order modeled. In combination with data-driven approaches such as modal decomposition or machine learning, DNS and LES constitute a very powerful tool for the identification of the dominant physical processes governing flow behavior under compressible and reactive conditions.

In parallel with developments in numerical simulation, modal decomposition methods have become the powerful tools of choice for interrogating the complicated, high-dimensional data generated by unsteady turbulent flows. The primary objective of these methods is to simplify the evident complexity of these flows by extracting coherent structures—spatiotemporal patterns—that dominate the dynamics. Among the various methods, Dynamic Mode Decomposition (DMD) has gained a lot of interest because it can break down flow fields into a sum of modes, each of which has a distinct frequency and a corresponding growth or decay rate. In contrast to conventional modal techniques like Proper Orthogonal Decomposition (POD), which emphasize the energy content of the system, Dynamic Mode Decomposition (DMD) offers a more nuanced perspective by directly identifying oscillatory behaviors from time-resolved datasets, independent of any physical model prerequisites.
The formulation of streaming variants of Dynamic Mode Decomposition (DMD), i.e., sDMD, has significantly improved the quality of the results by providing the reduced set of modes that are sufficient to describe the underlying dynamics of the system. It is especially beneficial for highly nonlinear and multiscale flow processes, for which classical decomposition methods give a large number of modes that are of little physical relevance. These modal techniques have already demonstrated their usefulness in a wide range of canonical flows, including incompressible and compressible jets, mixing layers, cylinder wakes, and boundary layers, by revealing leading instability mechanisms, transition dynamics, and large-scale flow organization \citep{jovanovic2014sparsity}.
Yet, their use for flows in which strong compressibility effects, shock waves, or heat releases are present—e.g., shock-impacted shear layers or reactions—is still relatively rare. In these flows, the occurrence of discontinuities and the interaction between acoustic, vortical, and thermal phenomena complicate matters further. The capacity of DMD and sDMD to de-tangle these interactions and pull out dynamically relevant structures provides a hopeful route to enhancing our knowledge of these regimes. Still more effort is needed to determine their efficacy and to modify these techniques to incorporate the particular characteristics of shock-turbulence and shock-flame interactions, thereby justifying the current research.
Although the majority of Dynamic Mode Decomposition studies are conducted in post-processing with time-resolved data acquired through simulations or experiments, typical implementations demand simultaneous access to the entire dataset. This is unfeasible when dealing with highly resolved turbulent flows because the memory demands for storing and processing such extensive data easily outstrip the available computational resources. Consequently, the usage of Dynamic Mode Decomposition (DMD) for very turbulent regimes has been relatively restricted. In an effort to overcome this restriction, several formulations have been suggested that enable the incremental calculation of DMD analysis concurrently with data acquisition. In particular, the streaming form of DMD (sDMD) presents an efficient algorithm that operates on data in a sequential manner with the need for only two time-consecutive snapshots at any one time. This method drastically lowers memory needs and makes real-time or in-place modal analysis feasible amidst high-fidelity simulation, e.g., Direct Numerical Simulation (DNS), and even in experimental diagnostics such as Particle Image Velocimetry (PIV). In addition, sDMD has also been found to converge to solutions that are in agreement with classical Dynamic Mode Decomposition (DMD), making it an efficient and scalable tool for the processing of large, time-varying datasets \citep{hemati2014dynamic}.

This work aims to fill this gap by performing a detailed modal analysis of a compressible turbulent mixing layer interacting with an oblique shock wave. Two configurations are considered: an inert shear layer and a reacting one with heat release due to hydrogen–air combustion. In both cases, the presence of the shock wave introduces additional mechanisms of mode selection and energy redistribution. To isolate the effects of shock interaction and combustion separately, simulations without the shock wave are also carried out for each regime. The comparative framework provides insights into how compressibility, shock modulation, and chemical reactions interact to shape the large-scale dynamics of the flow.
The analysis relies on the application of the streaming DMD technique to high-resolution datasets obtained from DNS. Particular attention is given to the identification of dominant frequency bands, mode structures, and the influence of heat release on modal growth and stability. The results provide a refined understanding of the physical mechanisms at play and offer perspectives for reduced-order modeling of such flows.

\section{Methodology}
\subsection{Mathematical Formulation and Computational Model}

The turbulent flow dynamics are investigated through direct numerical simulation (DNS) of the unsteady and compressible Navier–Stokes equations governing multicomponent reactive flows. These equations, formulated in conservative form, are given by:
\begin{equation}\label{eq:nseqs}
\left\{
\begin{aligned}
&\frac{\partial \rho}{\partial t} + \nabla \cdot (\rho \boldsymbol{u}) = 0, \\
&\frac{\partial (\rho \boldsymbol{u})}{\partial t} + \nabla \cdot (\rho \boldsymbol{u} \otimes \boldsymbol{u}) = \nabla \cdot \boldsymbol{\sigma}, \\
&\frac{\partial (\rho E_{t})}{\partial t} + \nabla \cdot (\rho \boldsymbol{u} E_{t}) = \nabla \cdot (\boldsymbol{\sigma} \cdot \boldsymbol{u} - \boldsymbol{q}), \\
&\frac{\partial (\rho Y_{\alpha})}{\partial t} + \nabla \cdot (\rho \boldsymbol{u} Y_{\alpha}) = - \nabla \cdot (\rho \boldsymbol{V}_{\alpha} Y_{\alpha}) + \rho \dot{\omega}_{\alpha}, \quad \alpha \in \mathscr{S} = \{1, \ldots, \mathcal{N}_{\text{sp}}\}.
\end{aligned}
\right.
\end{equation}
In these equations, $t$ represents time, $\boldsymbol{\nabla}$ is the spatial gradient operator, $\rho$ is the density, and $\boldsymbol{u}$ is the velocity vector. The total specific energy $E_{t}$ is defined as $e + \frac{1}{2} \boldsymbol{u} \cdot \boldsymbol{u}$, where $e$ denotes the internal energy. The species mass fractions are denoted by $Y_{\alpha}$, with corresponding diffusion velocities $\boldsymbol{V}_{\alpha}$ and chemical source terms $\dot{\omega}_{\alpha}$. The total number of species is given by $\mathcal{N}_{\text{sp}}$, and the heat flux vector is represented by $\boldsymbol{q}$. The system is closed using constitutive relations derived from kinetic theory. The thermodynamic state is described by the ideal gas equation of state $P = \rho R T / \mathcal{W}$, where $P$ is the pressure, $T$ the temperature, and $R$ the universal gas constant. The mixture molar mass $\mathcal{W}$ is calculated as $\mathcal{W}^{-1} = \sum_{\alpha=1}^{\mathcal{N}_{\text{sp}}} Y_{\alpha} / \mathcal{W}_{\alpha}$, based on the individual species molar masses $\mathcal{W}_{\alpha}$.
The transport fluxes are modeled as:
\begin{equation}\label{eq:closureseqs}
\left\{
\begin{aligned}
&\rho Y_{\alpha} \boldsymbol{V}_{\alpha} = - \sum_{\beta \in \mathscr{S}} \rho Y_{\alpha} D_{\alpha, \beta} \left( \boldsymbol{d}_{\beta} + \chi_{\beta} X_{\beta} \nabla (\ln T) \right),\\
&\boldsymbol{q} = \sum_{\alpha \in \mathscr{S}} \rho Y_{\alpha} \boldsymbol{V}_{\alpha} \left( h_{\alpha} + \frac{R T \chi_{\alpha}}{\mathcal{W}_{\alpha}} \right) - \lambda_{T} \nabla T,\\
&\boldsymbol{\sigma} = -P \boldsymbol{I} + \boldsymbol{\tau} = -P \boldsymbol{I} + \mu \left( \nabla \boldsymbol{u} + (\nabla \boldsymbol{u})^{\top} \right) + \lambda (\nabla \cdot \boldsymbol{u}) \boldsymbol{I},
\end{aligned}
\right.
\end{equation}
Here, $D_{\alpha, \beta}$ denotes the multicomponent diffusion coefficients, $\boldsymbol{d}_{\alpha}$ the diffusion driving forces, $\chi_{\alpha}$ the thermal diffusion ratios, $X_{\alpha}$ the mole fractions, $h_{\alpha}$ the species enthalpies, and $\lambda_T$ the thermal conductivity. The diffusion driving force is expressed as $\boldsymbol{d}_{\alpha} = \nabla X_{\alpha} + (X_{\alpha} - Y_{\alpha}) \nabla (\ln P)$. The shear and bulk viscosities are denoted by $\mu$ and $\lambda$, respectively. The heat flux encompasses three contributions, all of which are computed in this work: (i) heat conduction due to temperature gradients (Fourier term), (ii) enthalpy transport via mass diffusion, and (iii) the Dufour effect, which is the counterpart of the Soret effect. Similarly, the species diffusion flux includes: (i) Fickian diffusion modeled using the Hirschfelder-Curtiss formalism for multicomponent mixtures \cite{hirschfelder_molecular_1964}, (ii) pressure-gradient-induced diffusion (baro-diffusion) \cite{babkovskaia2011high}, and (iii) thermal diffusion (Soret effect) \cite{bird2006transport}. All transport properties are evaluated using the EGlib library \cite{ern1994multicomponent}, that is based on the kinetic theory \cite{kee2017chemically}. Finally, a detailed chemical kinetics mechanism is employed to account for $\mathcal{N}_{\text{reac}}$ elementary reaction steps involving $\mathcal{N}_{\text{sp}}$ chemical species. Each reaction is expressed in the general form
\begin{equation}
\sum_{\alpha=1}^{\mathcal{N}_{sp}} \nu_{\alpha,j}^\prime \mathcal{M}_\alpha 
\rightleftharpoons 
\sum_{\alpha=1}^{\mathcal{N}_{sp}} \nu_{\alpha,j}^{\prime\prime} \mathcal{M}_\alpha,
\quad j = 1, \ldots, \mathcal{N}_{\text{reac}},
\label{eq:elementary_reactions}
\end{equation}
where $\mathcal{M}_\alpha$ represents the chemical symbol of the $\alpha$th species, $\nu_{\alpha,j}^\prime$ and $\nu_{\alpha,j}^{\prime\prime}$ are the stoichiometric coefficients associated with the reactants and products of the $j$th reaction, respectively~\cite{kee1996chemkin}.  
The net production rate of species $\alpha$ due to chemical reactions is given by
\begin{equation}
\dot{\omega}_\alpha = \sum_{j=1}^{\mathcal{N}_{\text{reac}}} \left( \nu_{\alpha,j}^{\prime\prime} - \nu_{\alpha,j}^\prime \right) \left[k_{fj} \prod_{\alpha=1}^{\mathcal{N}_{sp}} [X_\alpha]^{\nu_{\alpha,j}^\prime}
- k_{rj} \prod_{\alpha=1}^{\mathcal{N}_{sp}} [X_\alpha]^{\nu_{\alpha,j}^{\prime\prime}}\right],
\end{equation}
where $[X_\alpha]$ denotes the molar concentration of species $\alpha$, and $k_{fj}$ and $k_{rj}$ are the forward and reverse rate constants of the $j$th elementary reaction, respectively. 
The forward and reverse rate constants are typically described by Arrhenius-type expressions of the form
\begin{equation}
k_{fj} = A_j T^{\beta_j} \exp\left( -\frac{E_{a,j}}{RT} \right),~k_{rj} = A_{rj} T^{\beta_{rj}} \exp\left( -\frac{E_{a,rj}}{RT} \right),
\end{equation}
where $A_j$ and $A_{rj}$ are the pre-exponential factors, $\beta_j$ and $\beta_{rj}$ are the temperature exponents, $E_{a,j}$ and $E_{a,rj}$ are the activation energies of the forward and reverse reactions, respectively. These empirical parameters are specific to each reaction and are obtained from detailed kinetic mechanisms.

All simulations are carried out using the \textit{Izem} solver. Inviscid fluxes are computed using a hybrid finite-difference framework that couples a non-dissipative eighth-order central differencing scheme—ideal for resolving fine-scale turbulence—with a shock-capturing seventh-order WENO method and Roe flux splitting in discontinuous regions. Shock zones are detected using an improved sensor based on normalized gradients of pressure and density \cite{adams_high-resolution_1996}.
Viscous and diffusive terms are discretized with eighth-order central differencing, while time advancement relies on a third-order Total Variation Diminishing (TVD) Runge–Kutta scheme. To address the stiffness arising from fast chemical kinetics, temporal integration of the source terms is performed using the CVODE solver from the Sundials suite \cite{hindmarsh2005sundials}, within an operator splitting framework similar to that described in \cite{ziegler2011adaptive}.
Thermodynamic properties, including species enthalpies and specific heat capacities, are modeled using seventh-degree polynomial fits as functions of temperature. These coefficients are obtained from the JANAF thermochemical tables to ensure accuracy and consistency in reactive simulations.
The high-order numerical framework implemented in \textit{Izem} has been extensively validated against benchmark experiments and DNS datasets for a variety of flow regimes. These include compressible homogeneous isotropic turbulence \cite{boukharfane2021triple}, shear layers \cite{boukharfane2021direct,boukharfane2021skewness}, underexpanded sonic jets \cite{baaziz2024large}, and supersonic crossflow–jet interactions \cite{boukharfane2022reacting}, demonstrating both its robustness and predictive capability.

\subsection{Problem statement and computational setup}
We study the interaction of an oblique shock wave with a spatially developing shear layer. The upper stream corresponds to the fuel inlet, i.e., a hydrogen-containing mixture, while the lower stream corresponds to vitiated air.
\begin{figure}[ht!]
\centering
\includegraphics[width=0.79\textwidth]{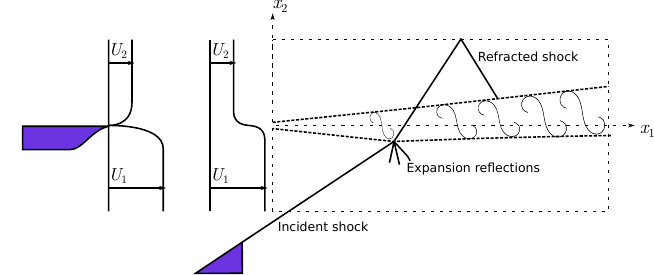}
\caption{Sketch of the two-dimensional shock–mixing layer interaction geometry.}
\label{fig:split}
\end{figure}
Figure~\ref{fig:split} shows a typical sketch of the computational geometry. The thermodynamic and velocity conditions imposed at the domain boundaries are summarized in Table~\ref{tab:theconfig}.

The flow initialization follows the approach used in \cite{ferrer2013etude}. Assuming equal specific heat capacity ratios in the two streams, the convective Mach number $\mathrm{Ma}_c$ and convective velocity $u_c$ are evaluated as 
\[
\mathrm{Ma}_{c} = \frac{u_1 - u_2}{c_{1} + c_{2}},~~u_c = \frac{u_1c_1 + u_2 c_2}{c_{1} + c_{2}}
\]
where $c_{1}$ and $c_{2}$ are the speeds of sound in the oxidizer (stream 1) and fuel (stream 2) inlets, respectively. For the present configuration, this yields $\mathrm{Ma}_{c} = 0.48$. The inlet velocities are $u_1 = 1634.0~\mathrm{m/s}$ for the oxidizer (bottom) stream and $u_2 = 973.0~\mathrm{m/s}$ for the fuel (top) stream.
An oblique shock wave, originating from the oxidizer stream at the bottom boundary, impinges on the mixing layer at an angle $\beta = 33^\circ$ (see Fig.~\ref{fig:split}). The computational domain spans a streamwise length of approximately $280$ initial vorticity thicknesses, and a crosswise extent of about $130$ initial vorticity thicknesses, where the initial vorticity thickness is set to $\delta_{\omega,0} = 1.44 \times 10^{-4}$~m. The grid is uniformly structured with $1640$ points in the streamwise direction and $750$ points in the transverse direction, ensuring adequate resolution for capturing both the shock structure and the shear-layer dynamics.
\begin{table}[h!]
\centering
\caption{Thermodynamic and velocity conditions at each boundary.}
\begin{tabular}{lccccc}
\toprule
\textbf{Zone} & $P$ (Pa) & $T$ (K) & $u_1$ (m/s) & $u_2$ (m/s) & $\mathrm{Ma}$ (-) \\
\midrule
Fuel      & 94232.25  & 545.0   & 973.0   & 0.0     & 1.60 \\
Oxidizer  & 94232.25  & 1475.0  & 1634.0  & 0.0     & 2.12 \\
Bottom    & 129951.00 & 1582.6  & 1526.3  & 165.7   & 1.93 \\
\bottomrule
\end{tabular}
\label{tab:theconfig}
\end{table}
The initial condition for the streamwise velocity is prescribed using a hyperbolic tangent profile to represent the shear layer interface, while the remaining velocity components are initialized to zero. The species mass fractions and density fields are defined consistently using the following general form:
\begin{equation}
\varphi\left(x_{1}, x_{2}\right) = \frac{\varphi_{1} + \varphi_{2}}{2} + \frac{\varphi_{1} - \varphi_{2}}{2} \tanh\left(\frac{2x_{2}}{\delta_{\omega,0}}\right),
\end{equation}
where $\varphi$ represents any scalar quantity, such as a species mass fraction or the streamwise velocity component. The initial velocity differential between the two streams, $\Delta u = u_1 - u_2$, together with the vorticity thickness $\delta_{\omega,0}$, yields a Reynolds number of $Re_{\delta_{\omega}} = 640$.
A key statistical quantity characterizing the evolution of the shear layer is its growth rate, often normalized by the streamwise coordinate. While several definitions exist, a commonly used expression relies on the vorticity thickness, defined as:
\begin{equation}
\delta_{\omega}(x_1) = \frac{u_1 - u_2}{\left.\partial \widetilde{u}_1 / \partial x_2\right|_{\max}},
\end{equation}
where $\widetilde{u}_1$ is the Favre-averaged streamwise velocity component.
The computational domain is bounded by various boundary conditions to reflect the physical characteristics of the flow. Supersonic inflow conditions are imposed using Dirichlet profiles for all conservative variables. The initial triggering of natural instabilities within the mixing layer is achieved by introducing perturbations defined as follows:
\begin{equation}
u_p = \varepsilon_1 \alpha u_c \exp \left( - \frac{\left(x_1-x_{1,0}\right)^2+\left(x_2-x_{2,0}\right)^2}{\Delta x_0^2} \right) \cos \left( \frac{2\pi \left(x_3-x_{3,0}\right) p_{\text{env}}}{L_3} + \varepsilon_2 \pi \right),
\label{eq:perturbation-3D}
\end{equation}
Here, $\varepsilon_1$ and $\varepsilon_2$ are random numbers varying between $-1$ and $+1$. The parameters $\alpha$ and $\Delta x_0$ represent the amplitude and length scale of the perturbation, respectively. Equation \eqref{eq:perturbation-3D} describes a sinusoidal variation along the spanwise direction, incorporating a random phase shift~\cite{bogey2000calcul}. The value of $p_{\text{env}}$ is set to introduce three complete periods of this sinusoidal function across the span. This perturbation is applied to both the transverse and spanwise velocity components, centered at $(x_{1,0},x_{2,0})=(4 \delta_{\omega,0},0)$.
A non-reflecting boundary condition is applied at the outflow to minimize wave reflections. A slip condition is enforced at the top wall, and the bottom boundary condition accounts for the incoming oblique shock through Rankine–Hugoniot relations, extended to handle multicomponent reactive mixtures~\cite{mitchell1982general}.
To trigger the transition to turbulence, a perturbation in the form of white noise with low amplitude is added to the transverse velocity component along the line $\left(x_1, x_2\right) = \left(4\delta_{\omega,0}, 0\right)$. Simulations are carried out using a Courant-Friedrichs-Lewy (CFL) number of $0.75$. Reactive simulations are carried out with the comprehensive detailed chemical kinetic mechanism reported by \citet{o2004comprehensive} with nine species (\ce{H2}, \ce{O2}, \ce{H2O}, \ce{H}, \ce{O}, \ce{OH}, \ce{HO2}, \ce{H2O2}, and \ce{N2}) and $21$ elementary reactions. The inlet stream compositions of fuel and oxidizer are determined based on chemical equilibrium considerations to enable autoignition in the computational model. While the shock mixing interaction is inherently a three-dimensional phenomenon, the use of a two-dimensional model in this study is justified by several considerations specific to the particular environment of internal flow and fuel-air mixing in a scramjet. First, most earlier studies of applicable internal flows for this particular application have also employed two-dimensional approximations \cite{kim2003numerical}. Second, \citet{sandham1990compressible} demonstrated that the mixing layer is two-dimensional in behavior when the convective Mach number is below some specific value (usually $0.62$). For the current simulation, the convective Mach number is found to be $0.48$, which is within the range where two-dimensional behavior prevails. Finally, \citet{javed2013model} proved that the mean pressure and velocity in similar configurations are largely described as two-dimensional. Based on these considerations and with regard to the focus of this study on the fundamental features of the shock-mixing interaction in internal flow, we believe that the selected two-dimensional approximation is suitable for our simulations and does not significantly affect the conclusions related to the consistency of the considered phenomenon.

\section{Modal decomposition of flow data}
\label{sec:dmd}
Before exploring the specific features and strengths of the streaming version of Dynamic Mode Decomposition (sDMD) \cite{hemati2014dynamic}, we first provide a brief overview of the fundamental principles behind DMD, with a focus on the classical algorithm based on Singular Value Decomposition (SVD) \cite{schmid2010dynamic}. This summary is restricted to the case of data sampled at regular time intervals. Readers interested in a broader or more rigorous treatment can consult \cite{kutz2016dynamic}.

Suppose a set of spatially distributed measurements is collected over time at a uniform sampling rate of $1/\Delta t$, resulting in $N$ snapshots. Each snapshot may consist of multi-dimensional data, which is reshaped into a column vector $\bm{x}_k \in \mathbb{R}^M$, where $k = 1, \dots, N$. Hence, the dataset forms a sequence $(\bm{x}_k)_{k=1}^{N}$ in $\mathbb{R}^M$. In the present context, $\bm{x}_k$ corresponds to the $k^{\text{th}}$ velocity field. For high-resolution three-dimensional flows, the size $M = 3 N_p^3$, where $N_p$ is the number of grid points, can become extremely large — an issue addressed in later sections.
The key assumption underlying DMD is the existence of a linear operator $\bm{A} \in \mathbb{R}^{M \times M}$ that approximately evolves the system in time over intervals of length $\Delta t$:
\begin{equation}
\label{eq:arnoldi}
\bm{x}_{k + 1} = \bm{A} \bm{x}_{k} + \bm{\varepsilon}_k \quad \text{for all} \quad k \in \{1, \dots, N-1\},
\end{equation}
where $\bm{\varepsilon}_k$ denotes a small residual error. This operator $\bm{A}$ is assumed time-invariant. The fidelity of this linear approximation depends on the system’s inherent timescales relative to the sampling interval, and more critically, on how well the nonlinear dynamics can be captured by a linear model.
In practical settings, the operator $\bm{A}$ is not known a priori and is instead estimated via a least-squares regression that minimizes the norm of the residual errors $\bm{\varepsilon}_k$ \cite{kutz2016dynamic}.
Since $\bm{A}$ encapsulates the system’s evolution in both space and time, its eigenvectors—termed ``dynamic modes'' or simply DMD modes—provide insight into the flow's coherent structures and are useful for developing reduced-order models. The next section outlines how these modes are extracted from the dataset $(\bm{x}_k)_{k=1}^{N}$, using an SVD-based method. This approach is preferred in practice for its numerical robustness and efficiency over more straightforward Krylov-subspace techniques. Further details can be found in the original work by \cite{schmid2010dynamic} and the in-depth discussion in \cite{kutz2016dynamic}.

\subsection{SVD-based DMD}  
\label{sec:svd-dmd}

To proceed, it is convenient to combine two data sequences of $N-1$ samples that are time-shifted by $\Delta t$, namely $(\bm{x}_k)_{\{k=1, \hdots, N-1\}}$ and $(\bm{x}_k)_{\{k=2, \hdots, N\}}$, into two matrices of dimensions $M \times (N-1)$:
\begin{align}
\label{eq:X1}
\bm{X} = \bm{X}_1^{N-1} = (x_{jk}) & : = \left( \bm{x}_1 \ \bm{x}_2 \ \cdots \ \bm{x}_{N-1} \right) \ , \\
\label{eq:X2}
\bm{Y} = \bm{X}_2^{N}   = (y_{jk}) & : = \left( \bm{x}_2 \ \bm{x}_3 \ \cdots \ \bm{x}_{N} \right) \ ,
\end{align}
where $j \in \{1, \hdots, M\}$ is the spatial index and $k$ the temporal one.
According to Eq.~\eqref{eq:arnoldi}, this leads to the model:
\beq
\label{eq:data-evol}
\bm{Y} = \bm{A} \bm{X} + \bm{R} \ ,
\eeq
where $\bm{R} = (\varepsilon_{jk})$ is a residual matrix. The matrix $\bm{A}$ that best approximates the linear mapping from $\bm{X}$ to $\bm{Y}$, in the least-squares sense, is given by:
\[
\bm{A} = \bm{Y}\bm{X}^+,
\]
where $\bm{X}^+$ is the pseudo-inverse of $\bm{X}$.
In practical scenarios—particularly in fluid dynamics—we usually have $M \gg N$, meaning the spatial dimension $M$ far exceeds the number of snapshots $N$. Consequently, the matrix $\bm{A} \in \mathbb{R}^{M \times M}$ is at most of rank $N-1$. This motivates the use of a reduced-order approximation by projecting $\bm{A}$ onto a lower-dimensional subspace spanned by $r$ POD modes, obtained through the compact SVD of $\bm{X}$:
\beq
\label{eq:svd}
\bm{X} =\bm{U}_{\bm{X}} \bm{\Sigma}_{\bm{X}} \bm{W}_{\bm{X}}^{T},
\eeq
where $T$ denotes the transpose. The truncation index $r$ is upper bounded by the rank of $\bm{X}$, at most $N-1$.
The matrix $\bm{U}_{\bm{X}} \in \mathbb{R}^{M \times r}$ contains the spatial structures (POD modes), $\bm{W}_{\bm{X}} \in \mathbb{R}^{(N-1) \times r}$ has orthogonal rows, and $\bm{\Sigma}_{\bm{X}} \in \mathbb{R}^{r \times r}$ is a diagonal matrix with the non-zero singular values.
By restricting $\bm{A}$ to act on the subspace spanned by the $r$ POD modes, we define the reduced matrix:
\beq
\label{eq:projection}
\bm{S} := \bm{U}_{\bm{X}}^{T} \bm{A} \bm{U}_{\bm{X}} = \bm{U}_{\bm{X}}^{T} \bm{Y} \bm{W}_{\bm{X}} \bm{\Sigma}_{\bm{X}}^{-1} \in \mathbb{R}^{r \times r} \ .
\eeq
This projection is interpreted in the least-squares sense, as it stems from computing the pseudo-inverse of $\bm{X}$ via its SVD and projecting $\bm{A}$ onto the POD basis using the orthogonality of $\bm{U}_{\bm{X}}$.
The eigenvalues of $\bm{S}$ approximate a subset of the non-zero eigenvalues of $\bm{A}$.
For practical implementation, the SVD-based DMD algorithm introduced by Schmid~\cite{schmid2010dynamic} proceeds by first collecting \( N \) equally spaced samples \( \left\{\bm{x}_{1}, \bm{x}_{2}, \ldots, \bm{x}_{N}\right\} \), where each sample \( \bm{x}_{j} \in \mathbb{R}^{M} \) for \( j \in \{1, \ldots, N\} \). From these snapshots, the data matrix \( \bm{X} \in \mathbb{R}^{M \times (N-1)} \) is constructed using the first \( (N-1) \) snapshots, as described in Eq.~\eqref{eq:X1}. A compact singular value decomposition (SVD) of \( \bm{X} \) is then performed (Eq.~\eqref{eq:svd}). Simultaneously, the matrix \( \bm{Y} \in \mathbb{R}^{M \times (N-1)} \) is constructed from the last \( (N-1) \) snapshots (Eq.~\eqref{eq:X2}), and the low-dimensional linear operator \( \bm{S} \) is computed according to Eq.~\eqref{eq:projection}. Once \( \bm{S} \) is obtained, its eigenvalues \( \lambda_k \) and eigenvectors \( \bm{v}_k \) are computed for \( k = 1, \ldots, r \), and the corresponding dynamic modes are finally recovered as 
\[
\psi_{k} = \bm{U}_{\bm{X}} \bm{v}_{k}.
\]
The flow field at time \blue{$t_s = s\Delta t$}, for integer $s$, can then be approximated using $N' \leq r$ dynamic modes:
\blue{
\beq
\label{eq:approx}
\bm{x}(t_s) = \bm{x}(s \Delta t) \approx \sum_{k=1}^{N'} b_{k} \lambda_k^s \psi_{k} \ ,
\eeq
}
where the coefficients $b_k$ are obtained by solving the equation at $s=0$ in the least-squares sense.
\blue{The logarithm of the eigenvalues $\lambda_k$ gives:}
\beq
\label{eq:eigenvalues}
\omega_{k} = \frac{\text{Im}(\ln(\lambda_{k}))}{\Delta t}, \quad \sigma_{k} = \frac{\text{Re}(\ln(\lambda_{k}))}{\Delta t}
\eeq
which are the angular frequency and temporal growth/decay rate of the $k^{\text{th}}$ DMD mode.
The approximation quality depends not only on the number of retained modes but also on the truncation index $r$. Several criteria exist to select $r$, such as the Optimal Singular Value Hard Threshold \cite{gavish2014optimal}, or sparsity-promoting algorithms \cite{jovanovic2014sparsity} that aim to retain a small number of dominant modes. \red{Kou and Zhang \cite{kou2017improved} also proposed an improved rank selection method based on the temporal evolution of each mode, leading to better convergence.}
Although DMD modes are often ranked by amplitude, this may not guarantee dynamical relevance. In the context of low-dimensional modeling, sparsity-promotion \cite{jovanovic2014sparsity} resolves this by minimizing the reconstruction error with an additional $L_1$ penalty, encouraging the use of fewer active modes.
As our interest lies in large-scale, statistically stationary flow features, we retain only modes with eigenvalues on or near the unit circle. These are sorted by increasing frequency, and only the lowest-frequency modes are kept. 
Note, however, that this selection strategy may be inadequate in flows with multiple dominant time scales. In such cases, temporal filtering or alternative selection methods may be more appropriate.

\subsection{Streaming DMD}

Traditional Dynamic Mode Decomposition (DMD) requires access to the complete dataset at once, which limits its applicability when dealing with large datasets due to memory constraints. This limitation is particularly problematic for data with high spatial resolution (large \( M \)) or long temporal sequences (large \( N \)). Streaming DMD, proposed by \citet{hemati2014dynamic}, overcomes this limitation by incrementally computing the POD-projected linear operator \( \bm{S} \) using only two snapshots at any time. This approach consists of two main components: a low-storage method for evaluating \( \bm{S} \), and a recursive update scheme based on Gram--Schmidt orthogonalization that progressively incorporates new data samples. The method begins by considering the standard snapshot matrices \( \bm{X} \) and \( \bm{Y} \) and reformulating the projected operator as:
\begin{equation}
\label{eq:sdmd}
\bm{S} = \bm{U}_{\bm{X}}^T \bm{Y} (\bm{U}_{\bm{X}}^T \bm{X})^+ 
= \bm{U}_{\bm{X}}^T \bm{U}_{\bm{Y}} \tilde{\bm{Y}} \tilde{\bm{X}}^+ 
= \bm{U}_{\bm{X}}^T \bm{U}_{\bm{Y}} \tilde{\bm{Y}} \tilde{\bm{X}}^T(\tilde{\bm{X}}\tilde{\bm{X}}^T)^+ 
= \bm{U}_{\bm{X}}^T \bm{U}_{\bm{Y}} \bm{H} \bm{G}_{\bm{X}}^+,
\end{equation}
where the matrices \( \tilde{\bm{X}} = \bm{U}_{\bm{X}}^T \bm{X} \) and \( \tilde{\bm{Y}} = \bm{U}_{\bm{Y}}^T \bm{Y} \) represent the projections of \( \bm{X} \) and \( \bm{Y} \) onto their respective POD bases. The matrices \( \bm{H} = \tilde{\bm{Y}} \tilde{\bm{X}}^T \) and \( \bm{G}_{\bm{X}} = \tilde{\bm{X}} \tilde{\bm{X}}^T \) are constructed from the projected data. This rearrangement not only simplifies the computation of \( \bm{S} \), but also improves memory efficiency, particularly when the projection ranks \( r_{\bm{X}} \) and \( r_{\bm{Y}} \) are smaller than the full spatial dimension \( M \), which is typical in many fluid mechanics applications. A crucial advantage of this formulation is that all terms on the right-hand side can be updated incrementally using a pair of samples \( \bm{x}_k \) and \( \bm{y}_k = \bm{x}_{k+1} \). As each new snapshot arrives, the orthogonal bases \( \bm{U}_{\bm{X}} \) and \( \bm{U}_{\bm{Y}} \) are updated using Gram--Schmidt orthogonalization, after which the current snapshots are projected onto their respective bases to yield \( \tilde{\bm{x}}_k \) and \( \tilde{\bm{y}}_k \). These are then used to update the matrices \( \bm{H} \) and \( \bm{G}_{\bm{X}} \) as
\begin{equation}
\bm{H} = \sum_{l = 1}^k \tilde{\bm{y}}_l \tilde{\bm{x}}_l^T, \qquad \bm{G}_{\bm{X}} = \sum_{l = 1}^k \tilde{\bm{x}}_l \tilde{\bm{x}}_l^T.
\end{equation}
Despite its advantages, the streaming approach has some limitations. The use of Gram--Schmidt for orthogonalization can affect numerical stability compared to alternatives like Householder reflections, which are not feasible here due to the method's sequential nature. Furthermore, noisy data may lead to full-rank matrices, reducing the computational benefits. In such cases, intermediate filtering strategies---described in detail in \cite{hemati2014dynamic}.---can be employed. Lastly, it is worth noting that the matrix \( \bm{G}_{\bm{X}} \) contains the squared singular values of \( \bm{X} \), which can be verified via singular value decomposition.

\section{Result and discussion}
\subsection{Flow characteristics}
\begin{figure}[ht!]
\centering
\includegraphics[width=0.89\textwidth]{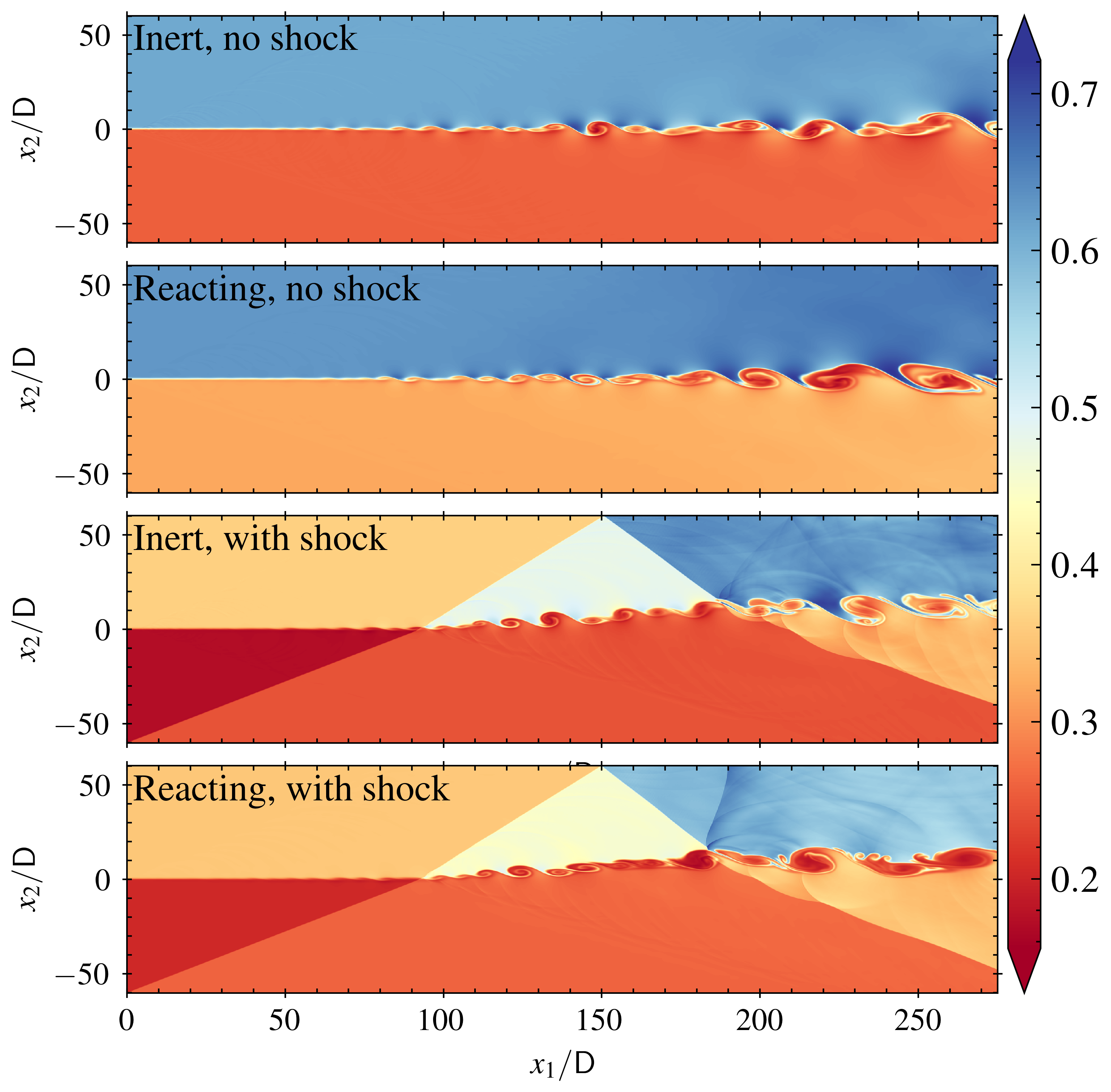}
\caption{Comparison of density distributions for the four different cases (inert, no shock; reacting, no shock; inert, with shock; reacting, with shock)}
\label{fig:inst_rho}
\end{figure}

The examination of the distributions of density across the four different scenarios is shown in Fig.~\ref{fig:inst_rho}.
The comparative examination of the instantaneous density fields for the four shear layer configurations reveals a notable structural similarity between the inert and reacting cases in the absence of an incident shock wave, where the development of Kelvin-Helmholtz instabilities leads to the formation and roll-up of coherent vortices with gradual density gradients. However, the introduction of an oblique shock wave significantly perturbs the density field in the shocked configurations (inert and reacting), inducing local compression visible through an increase in density, generating pressure waves and secondary instabilities that complicate the morphology of the vortices by deforming them and promoting their fragmentation, while also revealing the presence of reflected waves; although the fundamental vortical structures persist after the impact, their details are altered. Consistent with the observation that large-scale structures exhibit similarities between the inert and reacting cases, even with a shock, the influence of the chemical reaction on the instantaneous density field presented here is less direct. Nevertheless, as highlighted, the analysis of the temperature field (not shown) in the reacting case with a shock reveals the formation of isolated high-temperature regions as early as the first impact, where reactions initiate and intensify during the second reflected shock-shear layer interaction, leading to heat release that locally reduces the Mach number and can result in subsonic flow and the formation of an additional shock wave; while these thermal and compressibility effects induced by the reaction are not immediately apparent in the density field, they indirectly contribute to the dynamics and evolution of the mixing structures observed. In conclusion, the density field primarily illustrates the direct impact of the shock wave on compression and the complexification of structures, whereas the influence of the chemical reaction manifests more significantly through temperature variations and associated compressibility phenomena, which indirectly modulate the evolution of the mixing.

\begin{figure}[ht!]
\centering
\includegraphics[width=0.69\textwidth]{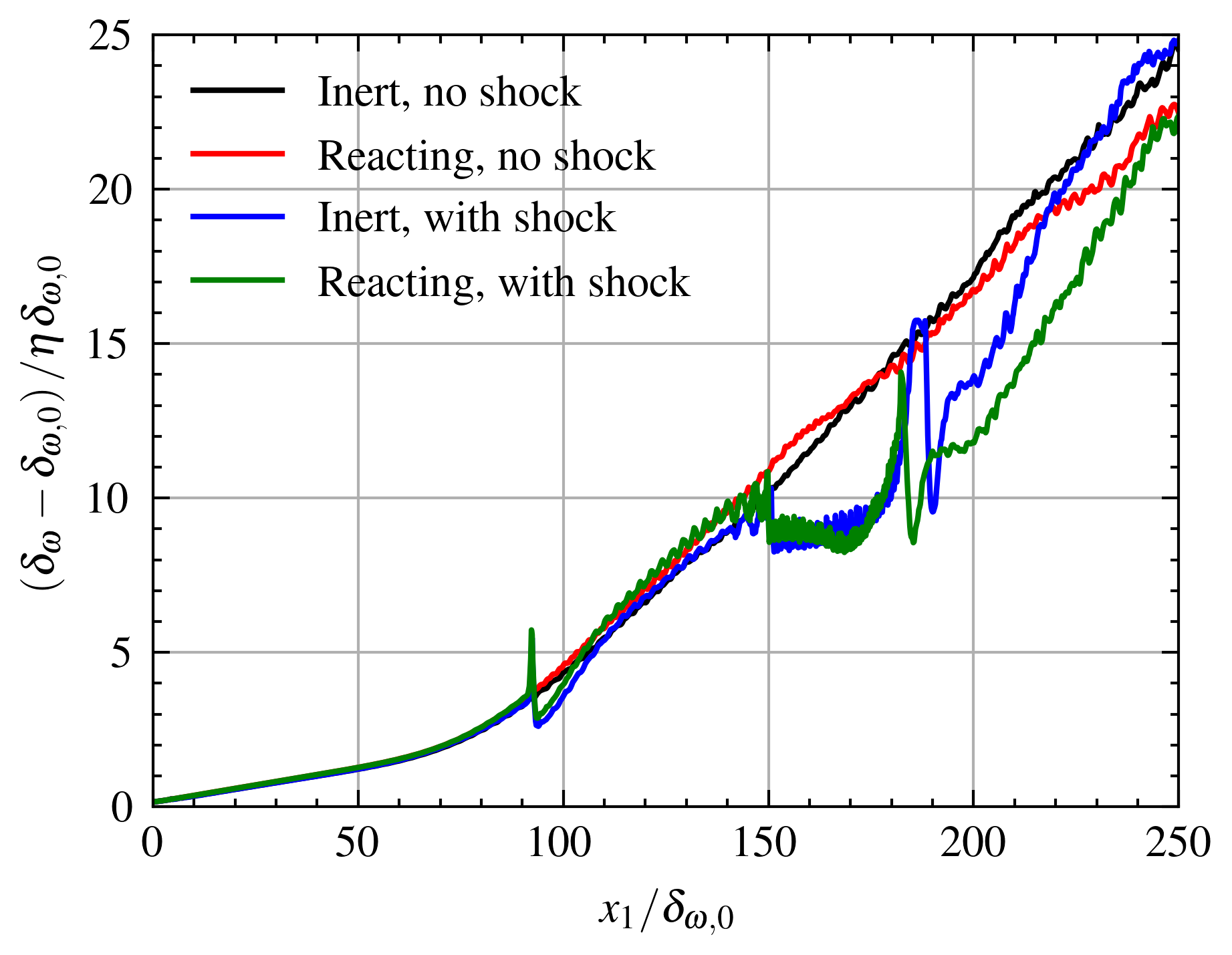}
\caption{Spatial evolution of the normalized vorticity thickness $(\delta_\omega - \delta_{\omega,0}) / \eta \delta_{\omega,0}$ as a function of the normalized streamwise coordinate $x_1 / \delta_{\omega,0}$ for the four simulated configurations (inert, no shock; reacting, no shock; inert, with shock; reacting, with shock).}
\label{fig:domega}
\end{figure}

The spatial development of the normalized vorticity thicknesses for the four cases under consideration is presented in Fig.~\ref{fig:domega}, highlighting characteristic dynamics that are affected by the presence of a shock wave and chemical reactions.
Initially, all the cases feature the same growth trend, indicating that the Kelvin-Helmholtz instability dominates the shear layer's development far from the shock's influence, with the reaction playing a negligible role at this scale. Nevertheless, the presence of a shock produces a sudden rise in the vorticity thickness growth rate following the initial interaction (at about $x_1 / \delta_{\omega,0} \approx 90-100$), which indicates the significant disturbance generated by the shock on the instability and mixing processes. After the initial interaction, the reactive case experiencing a shock has a marginally higher growth rate than the inert case for similar shock conditions, suggesting some benefit from the energy being released by the reaction. The non-physical oscillatory behavior, caused by reflections of shocks from the upper boundary of the computational domain, is seen at about $x_1 / \delta_{\omega,0} \approx 150$, which emphasizes the significance of boundary conditions when simulating shock phenomena. After the second interaction with the shock (around $x_1 / \delta_{\omega,0} \approx 180$), the influence of the heat release rate is much clearer, as supported by a wider difference between the reactive and inert cases in the shock case. This result points to a larger role of combustion in a flow already experiencing large perturbations. Against this, examples without a shock demonstrate a more gradual increase in the vorticity thickness with little effect of the reaction unless there are large external disturbances. Briefly, the effect of an oblique shock is to enhance greatly the turbulence and the thickening of the shear layer. Meanwhile, the chemical reaction, while relatively minor in comparison to begin with, shows a magnified effect when shock-generated disturbances are present, especially after several interactions. Additionally, numerical artifacts associated with shock reflections highlight the challenge of the numerical simulation of these phenomena.

\begin{figure}[ht!]
\centering
\subfloat[Inert, no shock]{\includegraphics[width=0.24\textwidth]{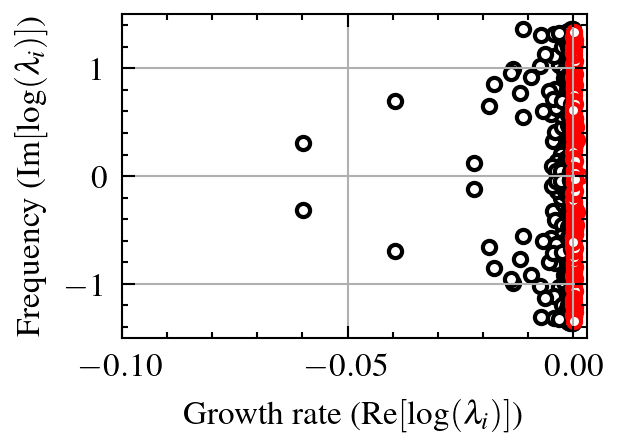}}
\subfloat[Reacting, no shock]{\includegraphics[width=0.24\textwidth]{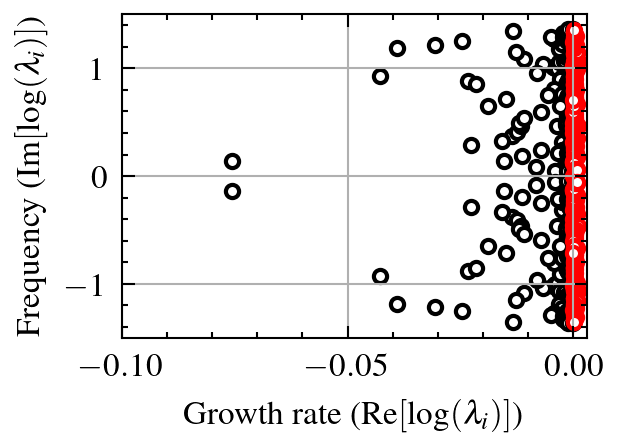}}
\subfloat[Inert, with shock]{\includegraphics[width=0.24\textwidth]{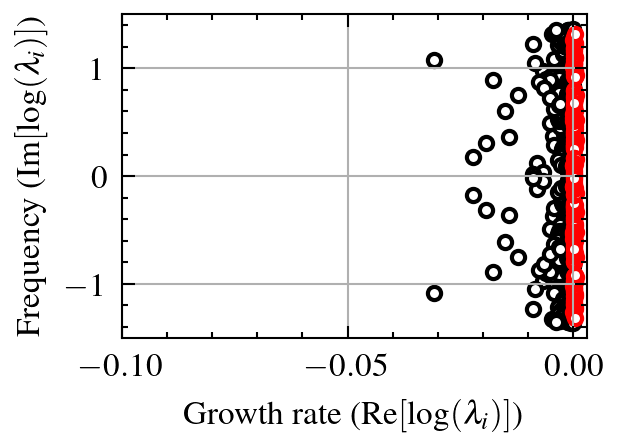}}
\subfloat[Reacting, with shock]{\includegraphics[width=0.24\textwidth]{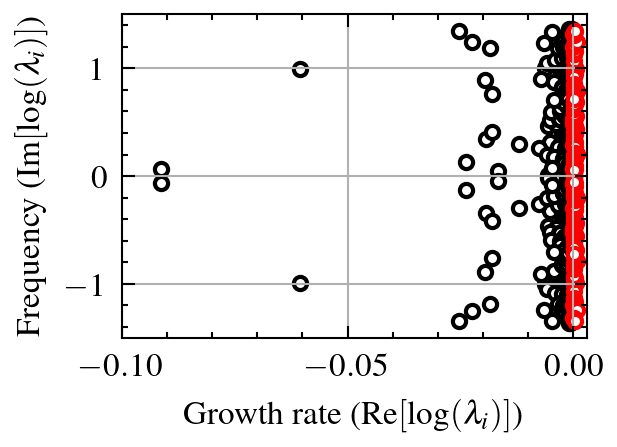}}
\caption{Comparison of the distribution of real and imaginary parts of the logarithmic eigenvalues.}
\label{fig:eig_modes}
\end{figure}

\subsection{Eigenvalue spectrum and energy contribution of DMD modes}

sDMD was performed using 1001 temporal snapshots of the unsteady flow fields. The resulting eigenvalue spectra in the complex plane for these four cases are presented in Fig. ~\ref{fig:eig_modes}.
In this representation, modes with non-positive real parts—indicating stable or neutrally stable behavior—are shown in black, whereas unstable modes with positive real parts are marked in red. In the absence of a shock wave, the inert mixing layer exhibits a predominantly stable behavior, with $85.90\%$ of the DMD modes being stable or neutrally stable and only $14.10\%$ identified as unstable, indicative of self-sustaining oscillations inherent to Kelvin-Helmholtz instability. However, the introduction of chemical reactions in the shock-free case leads to a notable destabilization, increasing the proportion of unstable modes to $23.70\%$ while stable modes decrease to $76.30\%$, suggesting that heat release from combustion introduces additional amplifying disturbances. The interaction with an oblique shock wave significantly alters the stability landscape. For the inert shock-mixing case, the percentage of unstable modes rises to $15.02\%$ ($84.98\%$ stable), indicating the shock's role in exciting flow instabilities, although the effect on the overall proportion of unstable modes is less pronounced than that of reaction alone in the shock-free scenario. Notably, the reacting shock-mixing interaction displays the highest fraction of unstable modes at $16.60\%$ ($83.40\%$ stable), highlighting a synergistic effect between shock-induced perturbations and heat release that further destabilizes the flow field. This quantitative analysis, complementing the qualitative observations from the eigenvalue spectra distributions, underscores the significant and distinct influences of both shock wave interaction and chemical reactions on the temporal stability of compressible shear layers. The increased prevalence of unstable modes in the reacting shock-impacted case likely contributes to the enhanced mixing and complex flow dynamics observed in such configurations.

\begin{figure}[ht!]
\centering
\subfloat[Inert, no shock]{\includegraphics[width=0.24\textwidth]{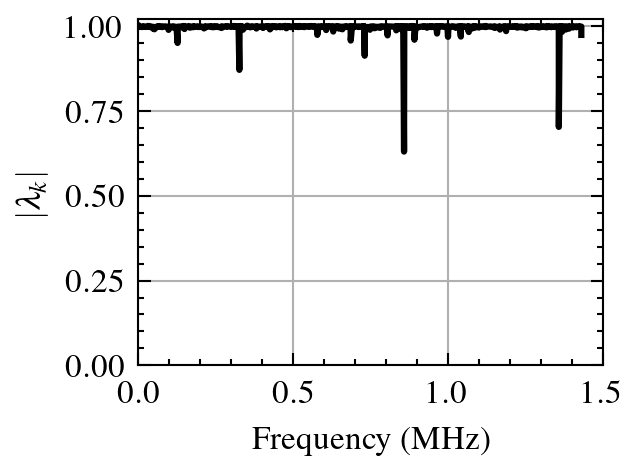}}
\subfloat[Reacting, no shock]{\includegraphics[width=0.24\textwidth]{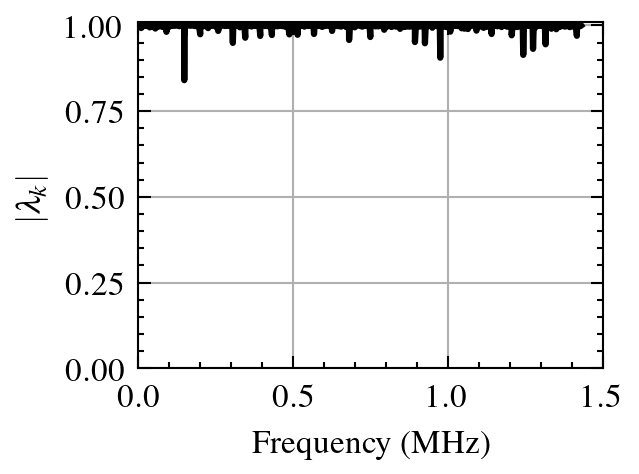}}
\subfloat[Inert, with shock]{\includegraphics[width=0.24\textwidth]{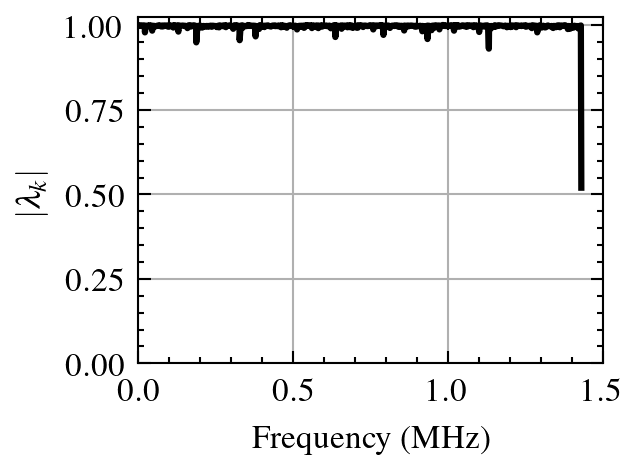}}
\subfloat[Reacting, with shock]{\includegraphics[width=0.24\textwidth]{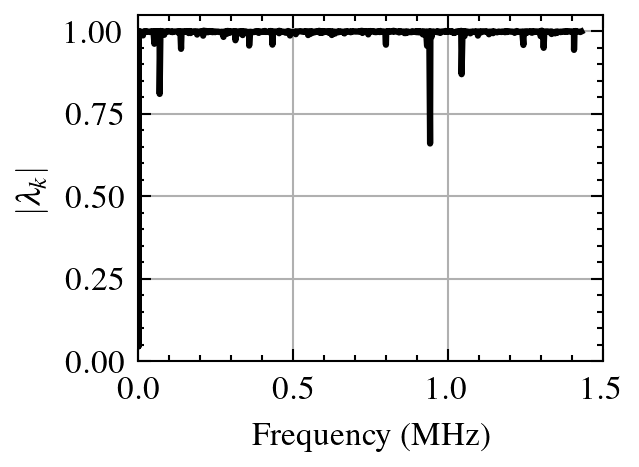}}
\caption{Comparison of the absolute eigenvalue magnitudes versus frequency.}
\label{fig:abs_val_eig_modes}
\end{figure}
Further analysis of the absolute values of the temporal mode eigenvalues ($|\lambda_k|$) at different frequencies provides a complementary perspective on the modal stability as shown in Fig.~\ref{fig:abs_val_eig_modes}. For all cases, the majority of modes exhibit $|\lambda_k|$ values close to 1, indicating a predominantly neutrally stable oscillatory behavior within the analyzed time window. However, deviations from unity reveal varying degrees of temporal growth and decay. In the inert shock-free case, $|\lambda_k|$ remains tightly clustered around 1, suggesting minimal amplification or damping. The reacting shock-free case shows a slightly broader spread, implying that reactions induce a wider range of temporal dynamics even without external forcing. The introduction of a shock in the inert case leads to more significant deviations from $|\lambda_k| = 1$, signifying the excitation of more unstable and stable modes. Notably, the reacting shock-mixing interaction displays the most extensive scattering of $|\lambda_k|$ values, with a clear presence of modes significantly greater than 1, reinforcing the synergistic destabilizing effect of shock and combustion on the temporal evolution of the flow structures. This analysis of the eigenvalue magnitudes corroborates the findings from the real parts of the logarithmic eigenvalues, providing a consistent picture of how shock waves and chemical reactions influence the temporal stability of the dominant dynamic modes in supersonic mixing layers.

\begin{figure}[ht!]
\centering
\subfloat[Inert, no shock]{\includegraphics[width=0.24\textwidth]{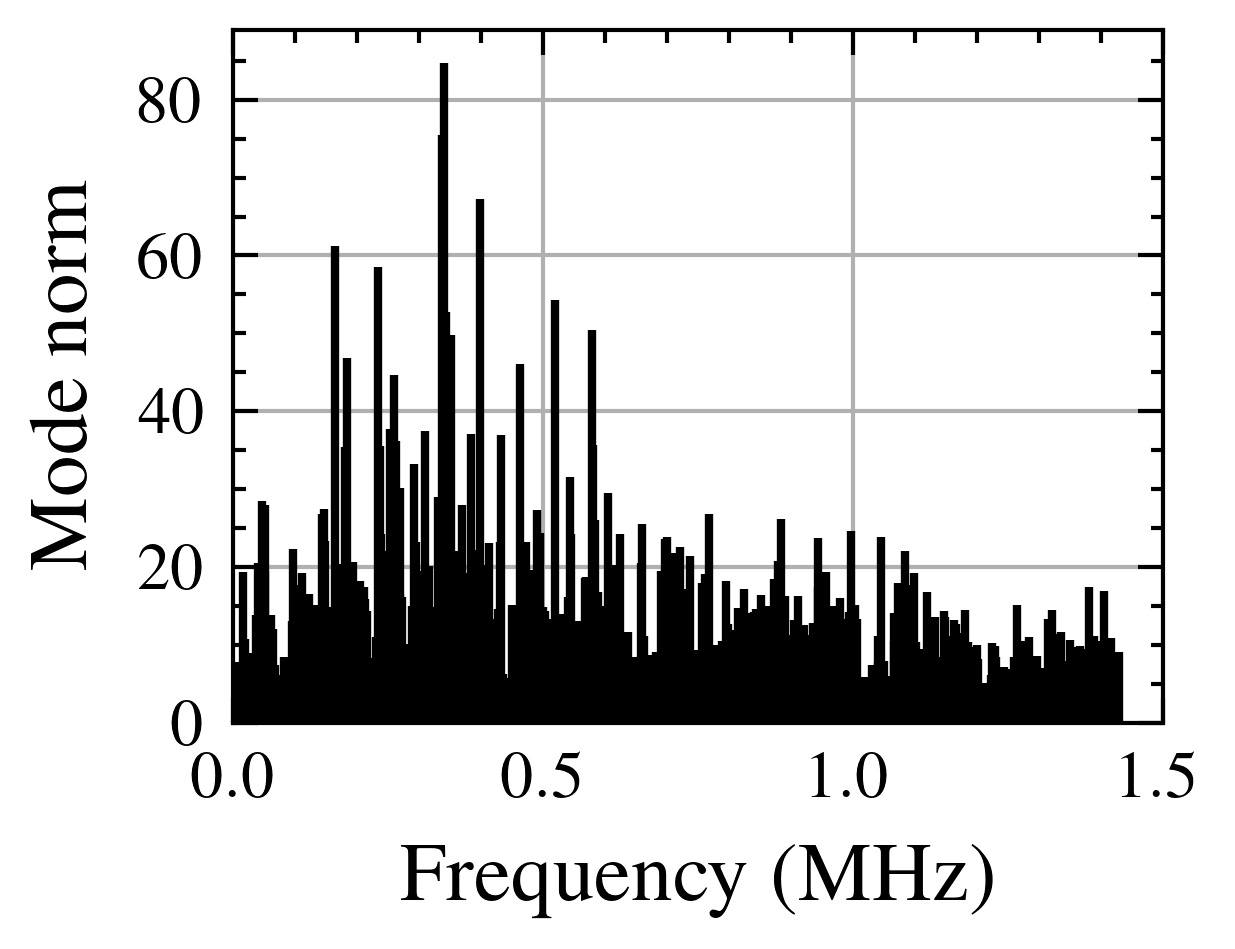}}
\subfloat[Reacting, no shock]{\includegraphics[width=0.24\textwidth]{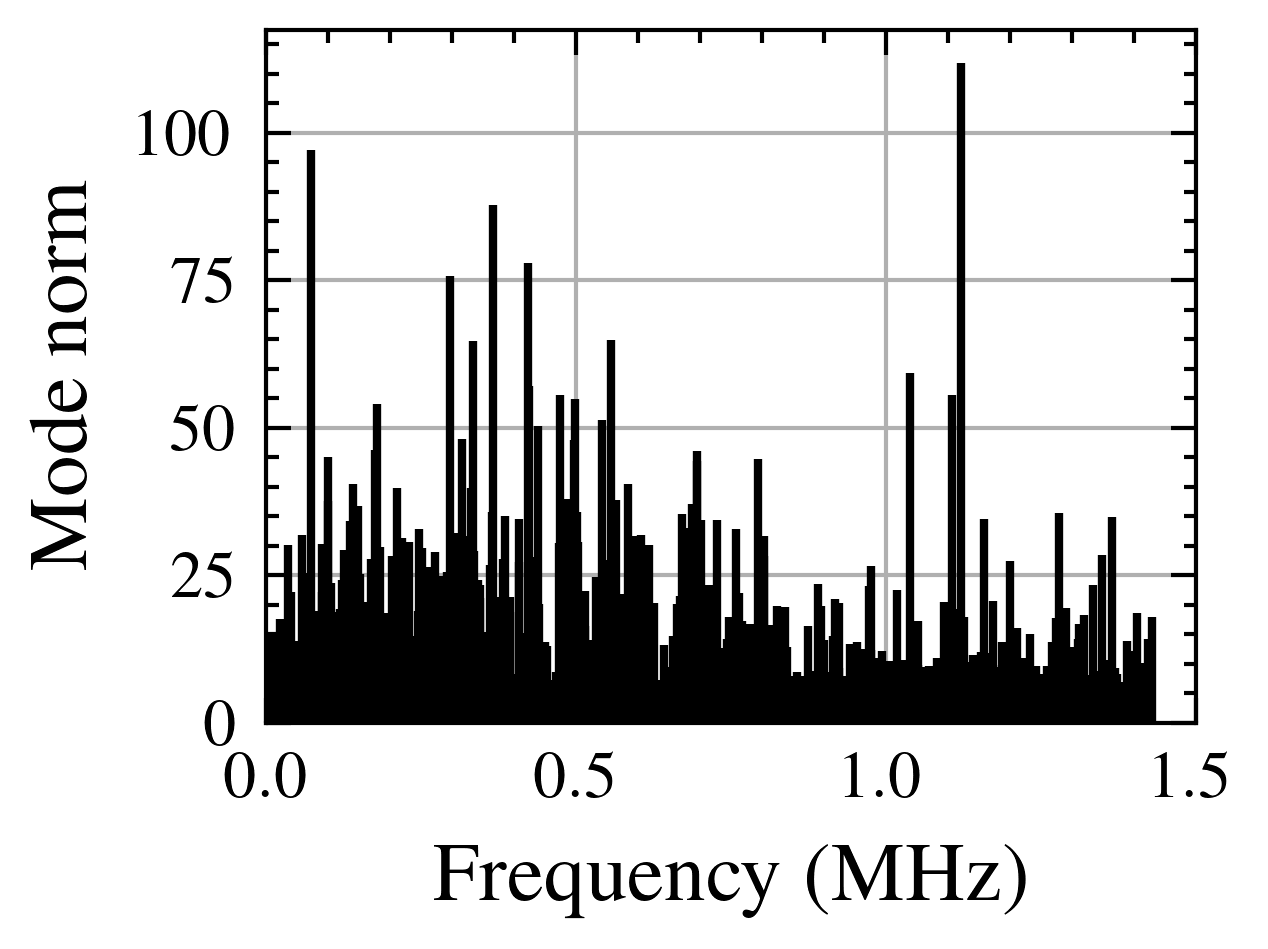}}
\subfloat[Inert, with shock]{\includegraphics[width=0.24\textwidth]{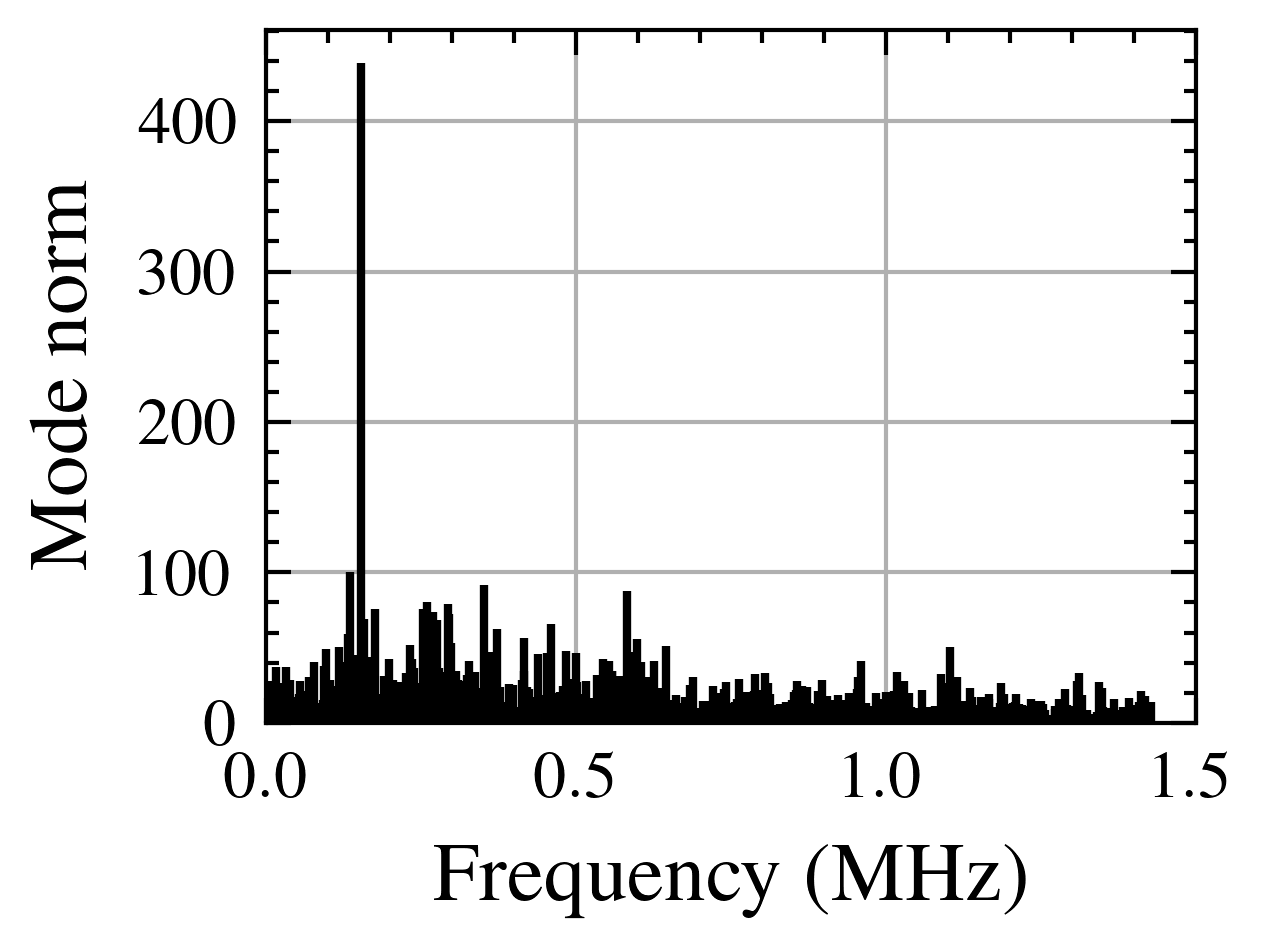}}
\subfloat[Reacting, with shock]{\includegraphics[width=0.24\textwidth]{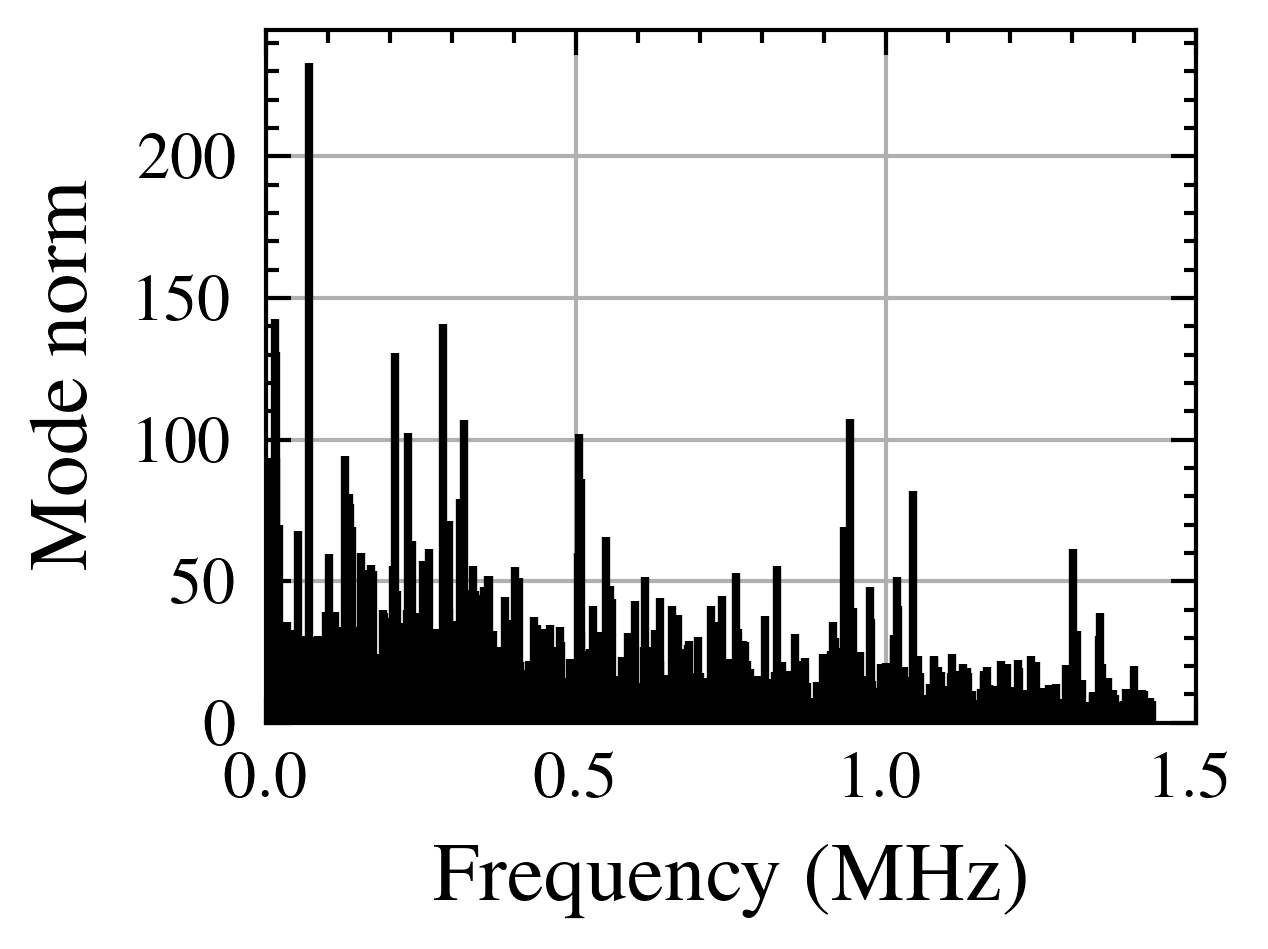}}
\caption{Comparison of the temporal mode analysis of the shear flow: 2-norms at different frequencies.}
\label{fig:normsfreq_modes}
\end{figure}

A comparative analysis of the temporal dynamics of inert and reacting shear layers, both with and without an impinging shock, based on the 2-norms of the DMD modes across a range of frequencies is presented in Fig.~\ref{fig:normsfreq_modes}. In the absence of shock, the energy of the modes is spread across a broad frequency spectrum, with dominant contributions below 0.5 MHz. Reacting mixing layers exhibit slightly higher mode norms than inert ones, indicating stronger coherent structures associated with combustion-induced perturbations. When the shock is introduced, the modal distributions are markedly altered. The inert case with shock interaction shows a pronounced peak at a low frequency ($\approx 0.05$ MHz), revealing the strong influence of shock-induced instability. This effect is also observed in the reacting case, though with more distributed energy over higher frequencies, reflecting a more complex and broadband dynamic response due to the combined effects of heat release and shock interaction.
A comparative analysis of the temporal dynamics is shown in Fig.~\ref{fig:normsfreq_modes}, using the 2-norms of the DMD modes across a range of frequencies.
In the absence of shock, the modal energy is distributed over a broad frequency spectrum, with dominant contributions below $0.5$~MHz. Reacting mixing layers exhibit slightly higher mode norms, indicating stronger coherent structures linked to combustion-induced perturbations. The introduction of a shock significantly modifies the modal distributions. In the inert case, a dominant peak appears at a low frequency ($\approx 0.05$~MHz), revealing the influence of shock-induced instability. This peak also appears in the reacting case but is accompanied by a broader distribution of energy across higher frequencies, reflecting the complex interplay between shock interaction and unsteady heat release. Across all cases, the dominant energy-containing modes are mostly temporally neutral, while the presence of the shock - especially under reacting conditions -- leads to a proliferation of low-energy modes with positive growth rates. These unstable modes likely contribute to the enhanced mixing and increased flow complexity observed in the reacting shock-interaction regime.

\begin{figure}[ht!]
\centering
\subfloat[Inert, no shock]{\includegraphics[width=0.24\textwidth]{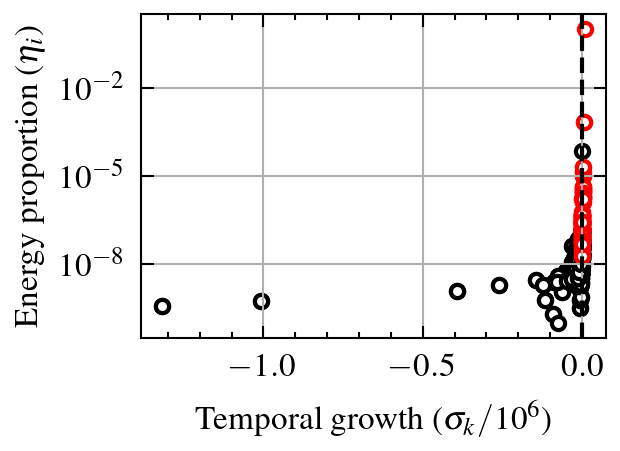}}
\subfloat[Reacting, no shock]{\includegraphics[width=0.24\textwidth]{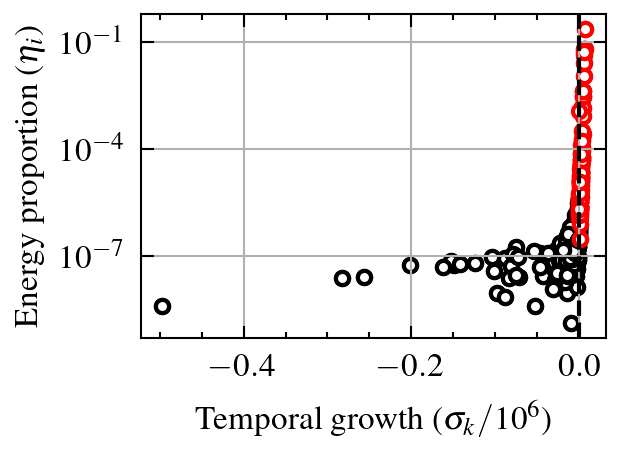}}
\subfloat[Inert, with shock]{\includegraphics[width=0.24\textwidth]{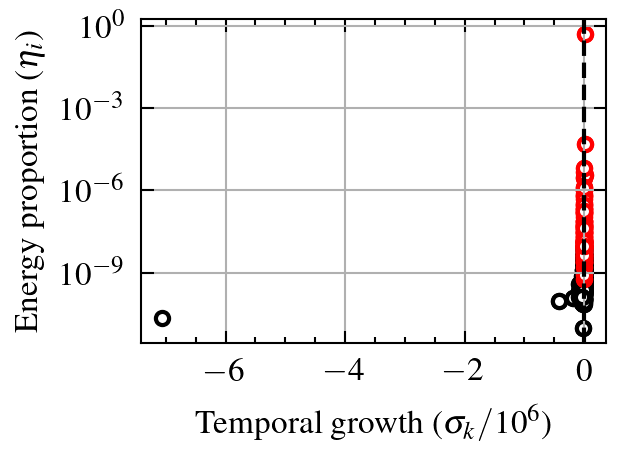 }}
\subfloat[Reacting, with shock]{\includegraphics[width=0.24\textwidth]{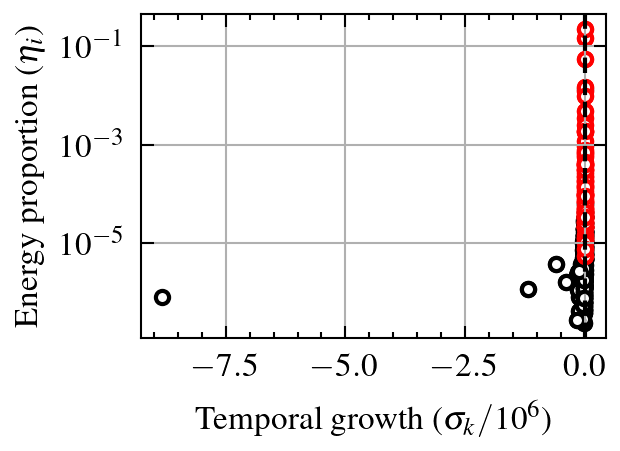 }}
\caption{Comparison of the distribution of mode energy proportion and growth rate.}
\label{fig:dmd_mode_energy_growth_modes}
\end{figure}

The energetic contribution of each DMD mode to the overall flow dynamics was quantified using the metric proposed by \citet{schmid2010dynamic}:
\begin{equation}
\mathcal{E}_i = \sum_{j=1}^{N} \left|c_{ij}\right| \left\| \boldsymbol{\lambda}_i \right\|_2,
\label{eq:mode_energy_scientific}
\end{equation}
where $c_{ij}$ is the $i$-th DMD complex temporal amplitude at the $j$-th snapshot. The total number of snapshots employed in the DMD analysis is $N = 1001$.
The fractional energy contribution of the $i$-th mode to the total energy of the system is then defined as:
\begin{equation}
\eta_i = \frac{\mathcal{E}_i}{\sum_{k=1}^{N} \mathcal{E}_k}.
\label{eq:energy_fraction_scientific}
\end{equation}
The relationship between the energetic significance and the temporal stability of the extracted modes is illustrated in Fig.~\ref{fig:dmd_mode_energy_growth_modes}, which presents the distribution of the energy fraction $\eta_i$ and the corresponding growth rate $\sigma_i$ for each DMD mode. 
Across all four cases, the majority of DMD modes exhibit negative growth rates, clustering towards the left side of the plots. This signifies that most of the identified modes are temporally stable and decay over time. However, a small subset of modes in each configuration shows growth rates close to or slightly above zero, indicating near-neutral stability or weak instability. Comparing the cases without shock to those with shock, we observe a noticeable difference in the distribution of modes. In the no-shock cases, the energetic modes (those with higher $\eta_i$ values) tend to have growth rates closer to zero, suggesting that the most energetically significant structures are either neutrally stable or only weakly decaying.
In contrast, the cases with shock interaction exhibit a broader spread in the growth rates of the energetic modes. While there are still energetically dominant modes with near-zero growth rates, there's a tendency for some of the more energetic modes to also possess more significant positive growth rates, particularly noticeable in the reacting case with shock. This implies that the shock interaction introduces or amplifies temporally unstable, yet energetically significant, flow structures.
Furthermore, the range of energy proportions spanned by the modes appears somewhat different across the cases. While all cases show a few dominant modes with relatively high energy fractions, the distribution of energy among the less energetic modes varies. The shock-impacted cases seem to have a larger number of modes with non-negligible energy contributions compared to the no-shock cases, hinting at a more complex and distributed energy landscape due to the shock-induced disturbances.
The comparison between the inert and reacting  cases reveals subtle differences. In the no-shock scenarios, the distribution of energy and growth rates appears qualitatively similar. However, with the introduction of the shock, the reacting case  seems to exhibit a slightly larger number of energetically significant modes with positive growth rates compared to the inert case with shock. This suggests that the interplay between the shock and the heat release from the chemical reactions might further destabilize certain energetic flow structures.

\subsection{DMD mode contour}

\begin{figure}[ht!]
\centering
\includegraphics[width=0.89\textwidth]{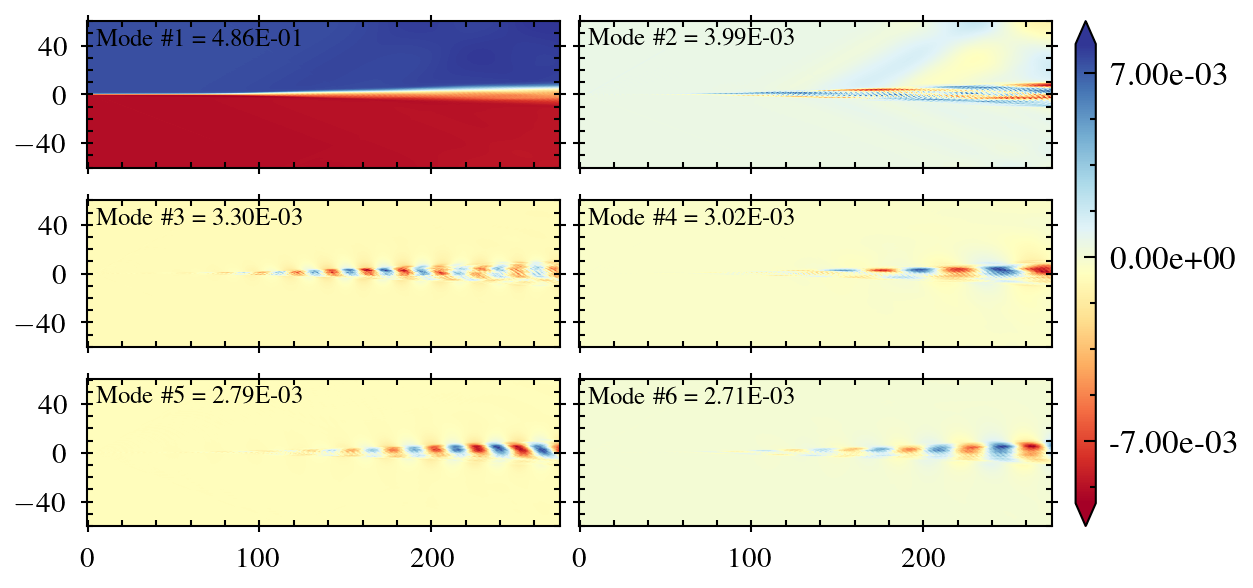}
\caption{Contours of the ﬁrst six orders DMD modes sorted by energy proportion for the inert shock-free mixing case}
\label{fig:dmdIsl}
\end{figure}
\begin{figure}[ht!]
\centering
\includegraphics[width=0.89\textwidth]{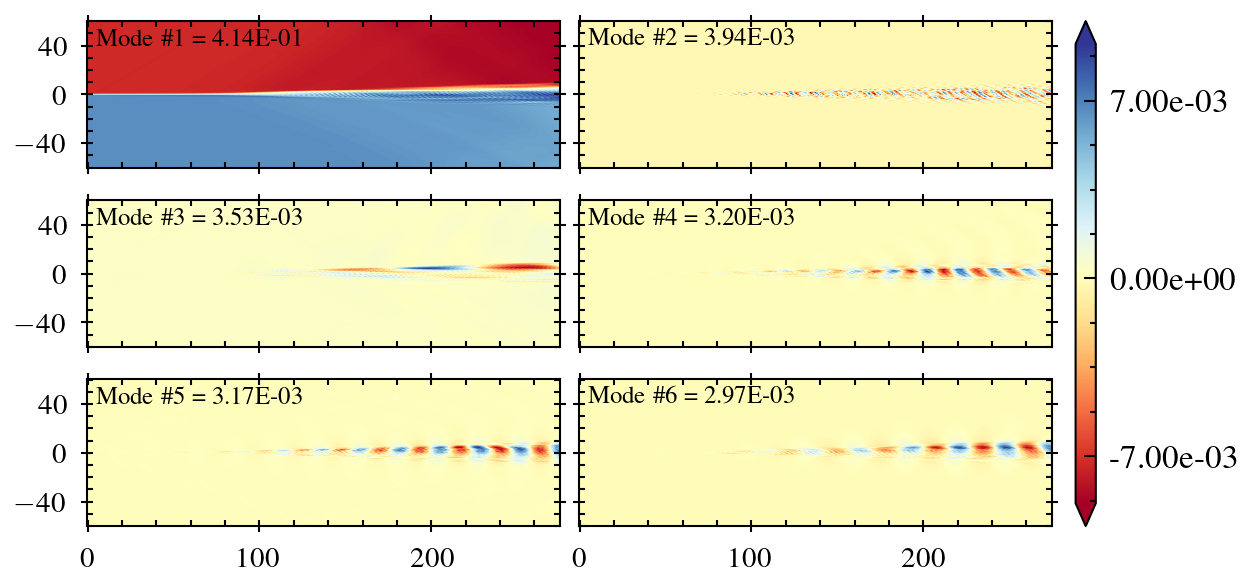}
\caption{Contours of the ﬁrst six orders DMD modes sorted by energy proportion for the reacting shock-free mixing case}
\label{fig:dmdRsl}
\end{figure}

The analysis of the first six energy-sorted DMD modes for the inert, shock-free shear layer in Fig.~\ref{fig:dmdIsl} reveals distinct spatial structures and their relative energetic importance.
The first mode, with a dominant energy proportion of 48.6\%, represents the time-averaged density field, capturing the mean state and containing nearly half of the flow's dynamic energy. The second mode, contributing 3.99\% of the energy, primarily reflects the pulsation of the vortex core associated with the primary Kelvin-Helmholtz instability, indicating a significantly less energetic but still crucial dynamic feature. The third mode, with 3.30\% energy proportion, mainly depicts the upper and lower boundaries of the mixing layer, highlighting regions of substantial density exchange and carrying a similar energy level to the vortex pulsation. The subsequent modes, 4th (3.02\%), 5th (2.79\%), and 6th (2.71\%), all exhibit a spatial stretch towards the lower region of the mixing layer, suggesting an influence of the lower Mach number on flow entrainment. These modes, while individually contributing a smaller fraction of the total energy, still represent coherent structures with distinct spatial characteristics. The rapid drop in energy proportion from Mode 1 to the subsequent modes underscores the dominance of the mean flow and the primary instability in the overall dynamics of the inert, shock-free mixing layer. The remaining modes capture progressively less energetic but still identifiable coherent structures and their spatial organization in the developing shear layer. The reacting case shown in Fig.~\ref{fig:dmdRsl} exhibits a lower energy proportion in the mean flow mode (Mode 1) and a slightly higher energy contribution in the mode representing the mixing layer boundaries (Mode 3). The energy associated with the vortex core pulsation (Mode 2) remains similar between the two. These shifts suggest that the energy landscape of the coherent structures is altered by the presence of chemical reactions, leading to a less dominant mean flow relative to the fluctuations and mixing processes compared to the inert case. The spatial structures themselves, while sharing qualitative similarities in representing the mean flow, vortex core dynamics, and mixing layer boundaries, likely exhibit more intricate details and potentially stronger gradients in the reacting case due to the influence of heat release and altered transport properties.

\begin{figure}[ht!]
\centering
\includegraphics[width=0.89\textwidth]{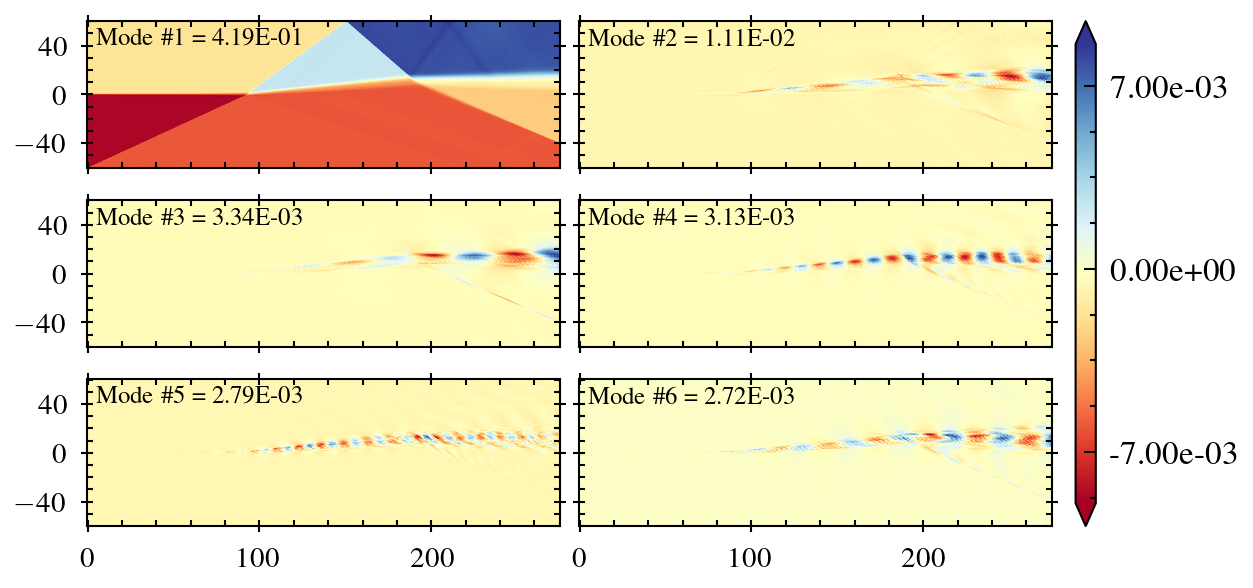}
\caption{Contours of the ﬁrst six orders DMD modes sorted by energy proportion for the inert shocked mixing case}
\label{fig:dmdIslSh}
\end{figure}

\begin{figure}[ht!]
\centering
\includegraphics[width=0.89\textwidth]{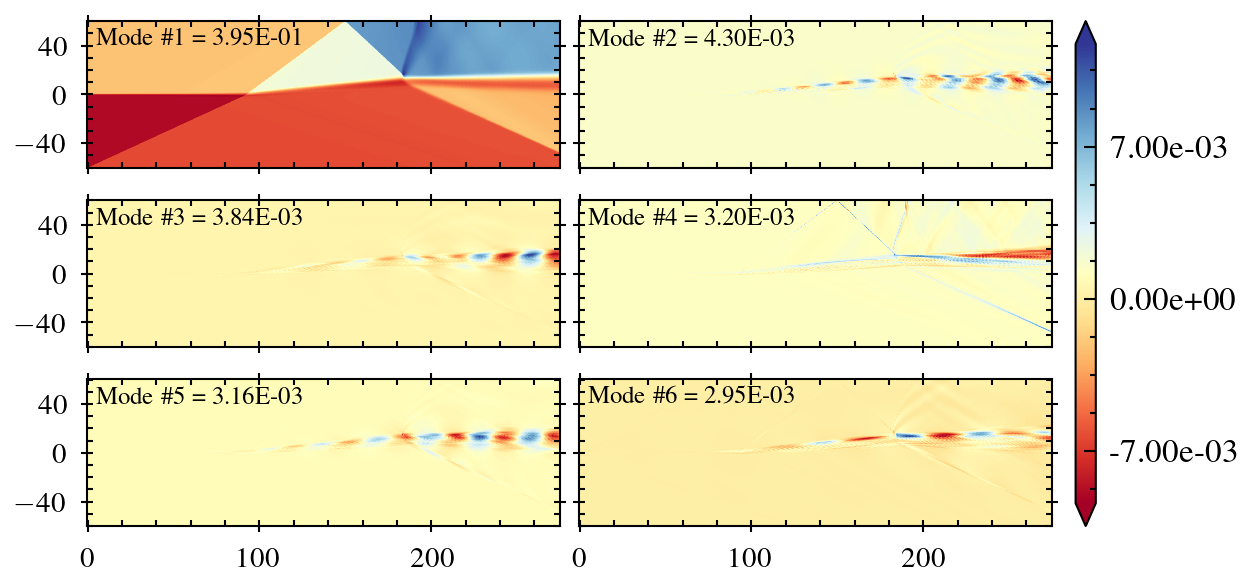}
\caption{Contours of the ﬁrst six orders DMD modes sorted by energy proportion for the reacting shocked mixing case}
\label{fig:dmdRslSh}
\end{figure}

The analysis of the first six energy-sorted DMD modes for the inert and reacting shear layers under shock interaction in Figs.~\ref{fig:dmdIslSh} and \ref{fig:dmdRslSh} reveals significant alterations in the energy distribution and spatial structures compared to the shock-free cases.
For the inert case with shock (energy proportions: $0.4193$, $0.01112$, $0.003337$, $0.003127$, $0.002785$, $0.002720$), the first mode, representing the dominant shocked structure, contains a substantial energy fraction. Notably, the second mode exhibits a considerably higher energy proportion than in the reacting shocked case ($0.3953$, $0.004302$, $0.003843$, $0.003201$, $0.003158$, $0.002947$), suggesting a stronger secondary shock-induced feature in the absence of reactions.
While the energy proportions of the higher-order modes are generally similar between the inert and reacting shocked cases, the reacting case tends to have slightly higher energy in modes 3 and 6. Spatially, the modes in the shocked cases are strongly influenced by the impinging shock, displaying compressed regions and wave-like patterns. In the reacting shocked case, these structures are further modified by the presence of flame fronts and heat release. The reduced energy in the second mode of the reacting shocked case, likely associated with shock-induced oscillations, and the altered energy distribution among higher-order modes highlight the complex interplay between shock waves and combustion in shaping the coherent structures of the supersonic mixing layer. Overall, the shock imposes a more dominant large-scale structure compared to the no-shock reacting case, and reactions significantly modify the energy partitioning among the shock-induced coherent structures.

\section{Conclusion}

This study has provided a detailed modal analysis of a compressible turbulent mixing layer interacting with an oblique shock wave, comparing inert and reacting configurations through Direct Numerical Simulations and the application of streaming Dynamic Mode Decomposition. Our findings reveal the distinct and interactive roles of shock impingement and chemical reactions in shaping the spatio-temporal dynamics of the shear layer.
In the absence of a shock, the temporal mode norms indicated a broad energy distribution across frequencies, with reacting layers exhibiting slightly enhanced coherent structures. The introduction of an oblique shock drastically altered the modal spectra, inducing pronounced low-frequency peaks indicative of strong shock-driven instabilities and a broader energy distribution towards higher frequencies, especially in the reacting case. Quantitative analysis of the eigenvalue spectra demonstrated a significant increase in the proportion of unstable modes upon shock interaction, further amplified by the presence of chemical reactions, suggesting a synergistic destabilizing effect.
The spatial evolution of the normalized vorticity thickness highlighted the shock's capacity to accelerate shear layer growth and the additional, albeit sometimes subtle, influence of heat release, particularly after multiple shock interactions. Modal decomposition of the flow field revealed the dominant coherent structures for each case. In the shock-free scenarios, the primary modes captured the mean flow and Kelvin-Helmholtz instability features, with reactions causing a redistribution of energy among these structures. The introduction of a shock led to the emergence of new, energetically significant modes directly related to the shock-induced perturbations. The reacting shocked case displayed the most complex modal structures, reflecting the intricate interplay between shock compression, turbulent straining, and chemical heat release.
Overall, this work underscores the critical role of oblique shock waves in fundamentally altering the dynamics and stability of compressible mixing layers, promoting instability and enhancing mixing. Chemical reactions, while having a discernible impact on their own, exhibit a more complex interaction with shock-induced disturbances, often amplifying the instability and modifying the energy distribution and spatial organization of the dominant coherent structures. These insights provide a valuable foundation for understanding and modeling high-speed reactive flows in applications such as scramjet engines, where shock-mixing interactions are inherent. Future work could explore the three-dimensional aspects of these interactions and investigate the sensitivity of these findings to variations in shock strength and fuel composition.

\subsection*{Acknowledgement}
This work was supported by OCP Group (Morocco). The authors gratefully acknowledge the support and computing resources from the African Supercomputing Center (ASCC) at UM6P (Morocco).
%

%
\bibliographystyle{plainnat}
\bibliography{main.bib}

\begin{thebibliography}{32}
\newcommand{\enquote}[1]{``#1''}
\providecommand{\natexlab}[1]{#1}
\providecommand{\url}[1]{\texttt{#1}}
\providecommand{\urlprefix}{URL }
\expandafter\ifx\csname urlstyle\endcsname\relax
  \providecommand{\doi}[1]{\discretionary{}{}{}https://doi.org/#1}\else
  \providecommand{\doi}[1]{\discretionary{}{}{}\urlstyle{rm}\url{https://doi.org/#1}}\fi

\bibitem[{Pirozzoli and Bernardini(2011)}]{pirozzoli2011turbulence}
Pirozzoli, S., and Bernardini, M., \enquote{Turbulence in supersonic boundary
  layers at moderate Reynolds number,} \emph{Journal of Fluid Mechanics}, Vol.
  688, 2011, pp. 120--168.

\bibitem[{Boukharfane et~al.(2018)Boukharfane, Bouali, and
  Mura}]{boukharfane2018evolution}
Boukharfane, R., Bouali, Z., and Mura, A., \enquote{Evolution of scalar and
  velocity dynamics in planar shock-turbulence interaction,} \emph{Shock
  Waves}, Vol.~28, No.~6, 2018, pp. 1117--1141.

\bibitem[{Mahesh(2013)}]{mahesh2013interaction}
Mahesh, K., \enquote{The interaction of jets with crossflow,} \emph{Annual
  Review of Fluid Mechanics}, Vol.~45, No.~1, 2013, pp. 379--407.

\bibitem[{Mathew et~al.(2008)Mathew, Mahle, and Friedrich}]{mathew2008effects}
Mathew, J., Mahle, I., and Friedrich, R., \enquote{Effects of compressibility
  and heat release on entrainment processes in mixing layers,} \emph{Journal of
  Turbulence}, , No.~9, 2008, p. N14.

\bibitem[{Fureby(2009)}]{fureby2009large}
Fureby, C., \enquote{Large eddy simulation modelling of combustion for
  propulsion applications,} \emph{Philosophical Transactions of the Royal
  Society A: Mathematical, Physical and Engineering Sciences}, Vol. 367, No.
  1899, 2009, pp. 2957--2969.

\bibitem[{Jovanovi{\'c} et~al.(2014)Jovanovi{\'c}, Schmid, and
  Nichols}]{jovanovic2014sparsity}
Jovanovi{\'c}, M.~R., Schmid, P.~J., and Nichols, J.~W.,
  \enquote{Sparsity-promoting dynamic mode decomposition,} \emph{Physics of
  Fluids}, Vol.~26, No.~2, 2014.

\bibitem[{Hemati et~al.(2014)Hemati, Williams, and Rowley}]{hemati2014dynamic}
Hemati, M.~S., Williams, M.~O., and Rowley, C.~W., \enquote{Dynamic mode
  decomposition for large and streaming datasets,} \emph{Physics of Fluids},
  Vol.~26, No.~11, 2014.

\bibitem[{Hirschfelder et~al.(1964)Hirschfelder, Curtiss, and
  Bird}]{hirschfelder_molecular_1964}
Hirschfelder, J., Curtiss, C., and Bird, R., \emph{Molecular theory of gases
  and liquids}, Structure of Matter Series, Wiley, 1964.

\bibitem[{Babkovskaia et~al.(2011)Babkovskaia, Haugen, and
  Brandenburg}]{babkovskaia2011high}
Babkovskaia, N., Haugen, N. E.~L., and Brandenburg, A., \enquote{A high-order
  public domain code for direct numerical simulations of turbulent combustion,}
  \emph{Journal of computational physics}, Vol. 230, No.~1, 2011, pp. 1--12.

\bibitem[{Bird et~al.(2006)Bird, Stewart, and Lightfoot}]{bird2006transport}
Bird, R.~B., Stewart, W.~E., and Lightfoot, E.~N., \enquote{Transport
  Phenomena,} , 2006.

\bibitem[{Ern and Giovangigli(1994)}]{ern1994multicomponent}
Ern, A., and Giovangigli, V., \emph{Multicomponent transport algorithms},
  Vol.~24, Springer Science \& Business Media, 1994.

\bibitem[{Kee et~al.(2017)Kee, Coltrin, Glarborg, and Zhu}]{kee2017chemically}
Kee, R.~J., Coltrin, M.~E., Glarborg, P., and Zhu, H., \emph{Chemically
  reacting flow: theory, modeling, and simulation}, John Wiley \& Sons, 2017.

\bibitem[{Kee et~al.(1996)Kee, Rupley, Meeks, and Miller}]{kee1996chemkin}
Kee, R.~J., Rupley, F.~M., Meeks, E., and Miller, J.~A., \enquote{{CHEMKIN-III:
  A Fortran chemical kinetics package for the analysis of gas-phase chemical
  and plasma kinetics},} \emph{Sandia national laboratories report
  SAND96-8216}, 1996.

\bibitem[{Adams and Shariff(1996)}]{adams_high-resolution_1996}
Adams, N.~A., and Shariff, K., \enquote{A {high}-{resolution} {hybrid}
  {compact}-{ENO} {scheme} for {shock}-{turbu\-lence} {interaction}
  {problems},} \emph{J. Comput. Phys}, Vol. 127, No.~1, 1996, pp. 27--51.

\bibitem[{Hindmarsh et~al.(2005)Hindmarsh, Brown, Grant, Lee, Serban, Shumaker,
  and Woodward}]{hindmarsh2005sundials}
Hindmarsh, A.~C., Brown, P.~N., Grant, K.~E., Lee, S.~L., Serban, R., Shumaker,
  D.~E., and Woodward, C.~S., \enquote{{SUNDIALS: Suite of nonlinear and
  differential/algebraic equation solvers},} \emph{ACM Transactions on
  Mathematical Software (TOMS)}, Vol.~31, No.~3, 2005, pp. 363--396.

\bibitem[{Ziegler et~al.(2011)Ziegler, Deiterding, Shepherd, and
  Pullin}]{ziegler2011adaptive}
Ziegler, J.~L., Deiterding, R., Shepherd, J.~E., and Pullin, D.~I., \enquote{An
  adaptive high-order hybrid scheme for compressive, viscous flows with
  detailed chemistry,} \emph{Journal of Computational Physics}, Vol. 230,
  No.~20, 2011, pp. 7598--7630.

\bibitem[{Boukharfane et~al.(2021{\natexlab{a}})Boukharfane, Er-raiy, Alzaben,
  and Parsani}]{boukharfane2021triple}
Boukharfane, R., Er-raiy, A., Alzaben, L., and Parsani, M., \enquote{Triple
  decomposition of velocity gradient tensor in compressible turbulence,}
  \emph{Fluids}, Vol.~6, No.~3, 2021{\natexlab{a}}, p.~98.

\bibitem[{Boukharfane et~al.(2021{\natexlab{b}})Boukharfane, Er-raiy, Elkarii,
  and Parsani}]{boukharfane2021direct}
Boukharfane, R., Er-raiy, A., Elkarii, M., and Parsani, M., \enquote{A direct
  numerical simulation study of skewed three-dimensional spatially evolving
  compressible mixing layer,} \emph{Phys. Fluids}, Vol.~33, No.~11,
  2021{\natexlab{b}}.

\bibitem[{Boukharfane et~al.(2021{\natexlab{c}})Boukharfane, Er-Raiy, Parsani,
  and Hadri}]{boukharfane2021skewness}
Boukharfane, R., Er-Raiy, A., Parsani, M., and Hadri, B., \enquote{Skewness
  effects on the turbulence structure in a high-speed compressible and
  multi-component inert mixing layers,} \emph{AIAA Aviation 2021 Forum},
  2021{\natexlab{c}}, p. 2915.

\bibitem[{Baaziz and Boukharfane(2024)}]{baaziz2024large}
Baaziz, S., and Boukharfane, R., \enquote{Large eddy simulation of highly
  underexpanded sonic jets from elliptical nozzles,} \emph{Physics of Fluids},
  Vol.~36, No.~10, 2024.

\bibitem[{Boukharfane et~al.(2022)Boukharfane, Techer, and
  Er-Raiy}]{boukharfane2022reacting}
Boukharfane, R., Techer, A., and Er-Raiy, A., \enquote{LES of reacting flow in
  a {H}ydrogen jet into supersonic crossflow combustor using a new turbulent
  combustion model,} \emph{International Journal of Aeronautical and Space
  Sciences}, Vol.~23, No.~1, 2022, pp. 115--128.

\bibitem[{Ferrer(2013)}]{ferrer2013etude}
Ferrer, P. J. M.~F., \enquote{{\'E}tude par simulation num{\'e}rique de
  l'auto-allumage en {\'e}coulement turbulent cisaill{\'e} supersonique,} Ph.D.
  thesis, ISAE-ENSMA-Poitiers, 2013.

\bibitem[{Bogey(2000)}]{bogey2000calcul}
Bogey, C., \enquote{Calcul direct du bruit a{\'e}rodynamique et validation de
  modeles acoustiques hybrides,} Ph.D. thesis, \'Ecole Centrale de Lyon, 2000.

\bibitem[{Mitchell and Kee(1982)}]{mitchell1982general}
Mitchell, R.~E., and Kee, R.~J., \enquote{A general-purpose computer code for
  predicting chemical-kinetic behavior behind incident and reflected shocks,}
  \emph{NASA STI/Recon Technical Report N}, Vol.~83, 1982, p. 27169.

\bibitem[{{\'O}~Conaire et~al.(2004){\'O}~Conaire, Curran, Simmie, Pitz, and
  Westbrook}]{o2004comprehensive}
{\'O}~Conaire, M., Curran, H.~J., Simmie, J., Pitz, W.~J., and Westbrook,
  C.~K., \enquote{A comprehensive modeling study of hydrogen oxidation,}
  \emph{International Journal of Chemical Kinetics}, Vol.~36, No.~11, 2004, pp.
  603--622.

\bibitem[{Kim et~al.(2003)Kim, Yoon, Jeung, Huh, and Choi}]{kim2003numerical}
Kim, J.-H., Yoon, Y., Jeung, I.-S., Huh, H., and Choi, J.-Y.,
  \enquote{Numerical study of mixing enhancement by shock waves in model
  scramjet engine,} \emph{AIAA journal}, Vol.~41, No.~6, 2003, pp. 1074--1080.

\bibitem[{Sandham and Reynolds(1990)}]{sandham1990compressible}
Sandham, N.~D., and Reynolds, W.~C., \enquote{Compressible mixing layer-linear
  theory and direct simulation,} \emph{AIAA journal}, Vol.~28, No.~4, 1990, pp.
  618--624.

\bibitem[{Javed et~al.(2013)Javed, Chakraborty, and Paul}]{javed2013model}
Javed, A., Chakraborty, D., and Paul, P.~J., \enquote{Model-free simulations
  for compressible mixing layer,} \emph{Proceedings of the Institution of
  Mechanical Engineers, Part G: Journal of Aerospace Engineering}, Vol. 227,
  No.~6, 2013, pp. 977--991.

\bibitem[{Schmid(2010)}]{schmid2010dynamic}
Schmid, P.~J., \enquote{Dynamic mode decomposition of numerical and
  experimental data,} \emph{Journal of Fluid Mechanics}, Vol. 656, 2010, pp.
  5--28.

\bibitem[{Kutz et~al.(2016)Kutz, Brunton, Brunton, and
  Proctor}]{kutz2016dynamic}
Kutz, J.~N., Brunton, S.~L., Brunton, B.~W., and Proctor, J.~L., \emph{Dynamic
  mode decomposition: data-driven modeling of complex systems}, SIAM, 2016.

\bibitem[{Gavish and Donoho(2014)}]{gavish2014optimal}
Gavish, M., and Donoho, D.~L., \enquote{The optimal hard threshold for singular
  values is $4/\sqrt{3}$,} \emph{IEEE Transactions on Information Theory},
  Vol.~60, No.~8, 2014, pp. 5040--5053.

\bibitem[{Kou and Zhang(2017)}]{kou2017improved}
Kou, J., and Zhang, W., \enquote{An improved criterion to select dominant modes
  from dynamic mode decomposition,} \emph{European Journal of
  Mechanics-B/Fluids}, Vol.~62, 2017, pp. 109--129.

\end{thebibliography}
\end{document}